\documentclass[10pt,preprint2]{aastex}

\usepackage{booktabs}
\usepackage{lscape}

\begin{document}

\title{EVOLUTIONARY TRACKS FOR BETELGEUSE}

\author{Michelle M. Dolan and Grant J. Mathews}
\affil{Center for Astrophysics, Department of Physics, University of Notre Dame \\ 
Notre Dame, IN 46656 \\
{\tt  velasnr@gmail.com \hspace{0.75em} gmathews@nd.edu \hspace{0.75em} }}

  \author{Doan Duc Lam}  
\affil{Department of Physics and Astronomy, Uppsala University, Uppsala, Sweden  \\
{\tt  lam.doan@physics.uu.se \hspace{0.75em}}}

  \author{Nguyen Quynh Lan}
\affil{Department of Physics, Hanoi National University of Education, Hanoi, Vietnam \\
{\tt   nquynhlan@hnue.edu.vn  \hspace{0.75em}}}

\author{Gregory J. Herczeg}
\affil{Kavli Institute for Astronomy and Astrophysics, Peking University, Beijing, China
 \\
{\tt gherczeg1@gmail.com}}

\author{David S. P. Dearborn}
\affil{Lawrence Livermore National Laboratory\\
Livermore, CA 94550 \\
{\tt ddearborn@llnl.gov}}

\begin{abstract} 
We have constructed a series of non-rotating quasi-hydrostatic  evolutionary models for the M2 Iab supergiant Betelgeuse ($\alpha~Orionis$).  Our models are constrained by multiple observed values for the   temperature, luminosity, surface composition and mass loss for this star, along with the parallax distance and high resolution imagery that determines its radius.  We have then applied our best-fit models to analyze the  observed variations in surface luminosity and the size of detected surface bright spots as the result of up-flowing convective material from regions of high temperature in the surface convective zone.  We also attempt to explain the intermittently observed periodic  variability in a simple radial linear adiabatic pulsation model.  Based upon the best fit to all observed data, we suggest a best progenitor mass estimate of $ 20 ^{+5}_{-3} M_\odot$ and a current age from the start of the zero-age main sequence  of $8.0 - 8.5$ Myr based upon the observed ejected mass while on the giant branch.

\end{abstract}


\section{Introduction}

The M2 Iab supergiant Betelgeuse ($\alpha~Orionis$) is an ideal laboratory to study advanced stages of stellar evolution.  It has the largest angular diameter of any star apart from the Sun and is one of the brightest M giants.   As such, it has been well studied.   Direct HST imagery exists for this star (Gilliland \& Dupree 1996; Lobel \& Dupree 2001; Dupree \& Stefanik 2013) as well as other high resolution indirect imagery (Balega et al.~1982;  Buscher et al.~1990; Marshall et al.~1992; Wilson et al.~1992, 1997;  Burns et al.~1997, Haubois et al.~2009; Townes et al.~2009; Ohnaka et al.~2009; 2013; Ohnaka 2014).  Both the light curve and the imagery indicate the appearance of intermittent bright spots associated with irregular variability in the star's luminosity and temperature.  The chromosphere has exhibited  (Dupree et al.~1987; Dempsey 2015) a periodic ($\sim$420 day) modulation in the optical and UV flux most likely associated with photospheric pulsations that  later became
substantially weaker, and  disappeared (Dupree \& Stefanik 2013).   The star  is classified (Samus et al.~2011) as a semi-regular variable with a SRC sub-classification with a period of 2335 $d$ ($6.39~y$).  

A shell of circumstellar material has also been detected around this star (Noriega-Crespo et al.~1997; Lobel \& Dupree 2001, Lobel 2003ab), and it appears to be losing mass at a rate of $\sim1$-$3$ $\times 10^{-6}$M$_\odot$ y$^{-1}$ (Knapp \& Morris 1985; Glassgold \& Huggins 1986; Bowers \& Knapp 1987; Skinner \& Whitmore 1987; Mauron 1990; Marshall et al.~1992; Young et al.~1993; Huggins et al.~1994; Mauron et al.~1995; Guilain \& Mauron 1996; Plez et al.~2002; Ryde et al.~2006; Harper et al.~2001, 2008; LeBertre et al.~2012; O'Gorman et al. 2015; Kervella, et al. 2016). 

 Isotopic CNO abundance data are also available  (Gautier et al.~1976; Harris \& Lambert 1984; Lambert et al.~1984) which suggest evidence of deep interior mixing.  On the other hand, this star exhibits a slow rotation velocity of $v \sin{i} = 5 $km s$^{-1}$ and an inclination of $i = 20^o$ (Kervella et al.~2009; Gilliland \& Dupree 1996) implying a rotation period of 8.4 y.  Hence, rotation may not significantly affect the interior structure at the present time, although it is likely to have affected  the main-sequence evolution (Meynet et al.~2013).  

These measurements have been complemented by the availability of high precision parallax measurements from the {\it Hipparcos} satellite (ESA 1997; Perryman et al.~1997; van Leeuwen 1997; Kovalevsky et al.1998), which have been revised (Harper et al.~2008).  The absolute luminosities and photospheric radii are now sufficiently well determined to warrant a new investigation of the constraints on models for this star.

In spite of this accumulated wealth of information there have been only a few attempts (e.g. Meynet et al.~2013) to apply a stellar evolution calculation in sufficient detail to explore the implications of these observed properties on models for the advanced evolution of this massive star.  Here we complement other such studies with an independent application of two quasi static stellar evolution codes from the pre-main sequence through the star's lifetime.  We have made a search to find the combinations of mass, mixing length, mass loss history, and age which best reproduce the observed radius, temperature, luminosity, current mass loss rate, and observed ejected mass for this star.  We then study the observed brightness variations, surface abundances, surface turbulent velocities and periodicity in the context of this model.  We find that both the observed intermittent periodicity and the hot-spot variability are  plausible outcomes of the surface convective properties of the model.

\section{Data}
\label{sec:data}
Over the years a great deal of data has accumulated for $\alpha$~Orionis (cf. Kervella, Le Bertre \& Perrin 2013; and refs therein).  Tables \ref{tab:1}-\ref{tab:5} summarize some of the observations and our adopted constraints as discussed below.  

\subsection{Distance}
In spite of its brightness, confusion in the proper motion, variability, asymmetry, and large angular diameter have together made the determination of the distance to this star difficult.  
A summary of the astrometric data is presented in Table \ref{tab:1}.  The first parallax distance measurements reported by {\it Hipparcos} ($\sim$131 pc) and {\it Tycho} ($\sim$54 pc) disagreed by more than a factor of two (ESA 1997).  This was well outside the range of quoted errors.  However, the more recent {\it VLA-Hipparcos} distance to Betelgeuse of $\sim197 \pm 45$ pc (Harper et al.~2008) has been derived from multi-wavelength observations.
This  is  the value adopted here as having the greatest accuracy and least distortion from the variability.

\subsection{Luminosity and Temperature}
Because of the variability of the star during observations, it is difficult to ascribe a mean value and  uncertainty to the visual magnitude, luminosity, and temperature.  The error bars associated with the quoted apparent visual magnitudes and temperatures in Table \ref{tab:2} are largely a measure of the intrinsic variability of the star.  They are therefore not a true measurement error.  Hence, to assign an uncertainty to the adopted mean visual magnitudes and temperatures  we simply take the un-weighted standard deviation of the various determinations of the mean value.  Since the luminosity depends upon distance, we simply adopt the luminosity of (Harper et al.~2008)  based upon the revised {\it Hipparcos} distance in that paper.   

\subsection{Angular Diameter}
Determination of the angular diameter $\Theta_{disk}$ and associated radius of this star from the observations are summarized in Table \ref{tab:3}.   This is also complicated by various factors.  Red supergiants like $\alpha$-Orionis are extended in radius.  As such, their surface gravity is  smaller than a main-sequence star like the Sun.  This results in extended atmospheres and large convective motion which produces asymmetry along with variability in the surface temperature and luminosity. 
In addition to  the random variability in surface luminosity and temperature and  intermittent periodic variability, the diameter of this star seems to change with time (Townes et al.~2009).  

Interferometric measurements can be used  to determine an effective uniform disk diameter.  However, measurements  made at one wavelength must be corrected for the variation in the optical depth with wavelength to yield an effective Rosseland mean radius to compare with the Rosseland radius computed in stellar models. 

One must also correct the inferred  disk for effects of limb darkening which tend to diminish the observed radius relative to the Rosseland mean radius.  This correction can be of order 10\% in visible wavelengths or only as little as $\sim 1$\% in the near infrared (Weiner 2000).  The presence of hot spots also tends to diminish the apparent radius (Weiner 2003) by as much as 15\%.  Moreover, the surrounding circumstellar envelope and dust also complicates the identification of the edge of the star.  Furthermore, measurements at different wavelengths lead to different results.  Infrared measurements over the past 15 years even seemed to indicate (Townes et al.~2009) that the radius of this star has been recently systematically decreasing [see, however,  Ohnaka et al.~(2013)].  

 Due to all of these complications one must exercise caution when using measurements of the angular diameter.   In Table \ref{tab:3}  we quote ( when available) the uniform disk Rossland mean radius corrected for limb darkening.  
  Quoted values in the literature are distributed into two distinct groups roughly depending upon wavelength.  One group $\lambda \sim 1~\mu m$) is centered around 44 mas (Cheng et al.~1986; Mozurkewich et al.~1991; Dyck et al.~1992; Perrin et al.~2004; Haubois et al.~2006, 2009) and while the other ($\lambda \sim 10~\mu m$) is centered around 57 mas (Balega et al.~1982; Buscher et al.~1990; Bester et al.~1996; Wilson et al.~1992, 1997; Burns et al.~1997; Tuthill et al.~1997; Weiner et al.~2000).  
  
  Photospheric radii  must be corrected for limb darkening and wavelength.  These corrections are the smallest for the 11 $\mu$m measurements (Wiener et al.~2000).  Hence, we adopt a weighted average of the 11.15 $\mu$m measurements to obtain $55.6 \pm0.04$ mas which would lead  to a photospheric Rosseland mean radius of  $56.2 \pm 0.04$ mas.  
  
  However, it is argued quite persuasively in Perrin et al.~(2004) that the discrepancy between the lower and higher radii could be accounted for in a unified model that includes the possibility of a warm molecular layer around the star, consistent with that also observed in Mira variables.  Applying this correction  to the 11.15 $\mu$m data leads to a 75\% correction (Perrin et al.~2004).  This reduces the corrected angular diameter to $41.9 \pm 0.04$ mas.  We adopt this as the best  means to deduce a radius to compare with stellar models.  This adopted  angular diameter, combined with the VLA-Hipparcos distance and uncertainty of $197 \pm 45$ pc, yields a radius of $887\pm 203$ R$_\odot$.  These parameters, along with the observed CNO abundances, lack of s-process abundances (Lundqvist \& Wahlgren 2005), and the ejected mass allow for constraints on stellar  models as described below.
 
 \subsection{Surface Composition}
The surface elemental and isotopic abundances adopted in this study are summarized in Table \ref{tab:4}. Surface elemental C, N, and O abundances for Betelgeuse abundances have been measured by  Lambert et al.~(1984), while $^{12}$C/$^{13}$C, $^{16}$O/$^{17}$O, and $^{16}$O/$^{18}$C isotopic ratios have been reported in Harris \& Lambert (1984).
As noted in those papers and in the calculations reported here, relative to the Sun, Betelgeuse has an enhanced nitrogen abundance,  low carbon abundance, and  low $^{12}$C/$^{13}$C ratio.  This is consistent with material that has been mixed to the surface as a result of the first dredge-up phase.  As we shall see, this places a constraint on the location of this star on its evolutionary track, i.e. that it must have passed the base of the red giant branch and is ascending the red supergiant phase.

However, to compare the observed abundances  with our computed models, some discussion is in order.
 To begin with, Lambert et al.~(1984) report [Fe/H] $= +0.1$ for this star (where $[X] \equiv \log{(X/X_\odot)}$.  We adopt  the Anders \& Grevesse (1989) proto-Solar values $X_\odot,Y_\odot,Z _\odot= 0.71,0.27, 0.020$, and assuming that  [Fe/H] is representative of metallicity, then $[Z]  = +0.1$.  This implies $Z = 0.024$ for this star.  Since the helium mass fraction correlates with metallicity $\Delta Y/\Delta Z = 3$ (Aver et al.~2013), this implies a helium mass fraction of $Y = 0.28$, from which one can infer $X = (1 - Y - Z) = 0.70$.  Hence, we adopt this composition. We note however, that  newer photospheric abundances (Asplund et al.~2009) would imply a lower metallicity than that adopted here.  
  
  Regarding the C, N, and O abundances, Lambert et al.~(1984) reported abundances relative to hydrogen:  $\epsilon$(C), $\epsilon$(N), and $\epsilon$(O), where $\epsilon(Z) \equiv \log(N(Z)/N(H) + 12$ as given in Table \ref{tab:4}.   They note that the inferred abundances depend upon the measured value of $T_{eff} = 3800$ K and the assumed value of $\log{g} = 0.0 \pm 0.3$ used in their model atmosphere.  Although the temperature  is different  from the value  adopted  in the present work,  this star shows considerable variability and  this was undoubtedly the appropriate temperature  during their observing epoch.  Hence, there is no need to correct their abundances for temperature.  
  
  Values of $\epsilon_i$ for individual isotopes are straightforward to evaluate from the isotope ratios given in Table \ref{tab:4}.  The isotopic mass fractions $X_i$  and estimated  uncertainties are then determined from our adopted hydrogen mass fraction $X_H$, i.e.
  \begin{equation}
  \log{X_i} =  \epsilon_{i } - 12 + \log{( X_H A)} ~~,
  \end{equation}
  where A is the atomic mass number.

\subsection{Mass loss and variability}
In Table \ref{tab:5}  we summarize variability data concerning mass loss, surface brightness features, and surface turbulent velocities.  As in Table \ref{tab:1}, the adopted values of these quantities represent a range of possible systematic errors plus the intrinsic variability of the star.  

Observations and models have deduced   
(Noriego-Crespo et al.~1997; Ueta et al.~2008; Mohamed, Mackey \& Langer 2012) that the combination of mass ejection and the supersonic motion of this star  relative to the local interstellar medium has led to the formation of  a bow  shock pointing along the direction of motion.   an estimate of the total ejected mass for this star can be deduced from infrared observations of the surrounding dust.  

Noriega-Crespo et al.~(1997) first  analyzed high resolution {\it IRAS} images at 60 and 100 $\mu m$ which indicate the presence of a shell bow shock.  They deduce a mass in the shell of 
\begin{equation}
M_{shell} = 0.042 ~{\rm M}_\odot~\biggl({F_{60} \over 135~{\rm Jy}}\biggr)\biggl({D \over 200 ~{\rm pc}}\biggr)~~,
\end{equation}
where $F_{60}$ is the  measured flux from the shell at 60 $\mu m$ which they determine to be $110 \pm 21$ Jy.  

For our adopted distance of 197 $\pm 45$  pc, the mass in this shell is then $0.034 \pm 0.016$ M$_\odot$  in  the immediate vicinity of Betelgeuse.  From this, a current mass-loss rate of  around $3 \pm 1 \times 10^{-6}$ M$_\odot$ $y^{-1}$  (Harper et al.~2001) can be deduced.   

Moreover, it is now recognized that there are multiple bows shocks (Mackey et al.~2012; 2013) indicating that the mass loss from this star is  episodic (Decin et al.~2012) rather than continuous.
 An additional detached shell  of neutral hydrogen has also been detected (Le Bertre et al.~2012) that extends out to 0.24 pc.  If one considers this shell,  then the total ejected mass increases to 0.086 M$_\odot$,  but the average  mass-loss rate reduces to $1.2 \times 10^{-6}$ M$_\odot$ $y^{-1}$ for the past 
$8 \times 10^4$ y.   For our purpose we will adopt a  mass loss rate of $2 \pm 1 \times 10^{-6} $ M$_\odot$ $y^{-1}$ encompassing both the immediate burst rate of Harper et al.~(2001)  and the average of Le Bertre et al.~(2012).  

On the other hand, one cannot be sure how much of the material in the bow shock is interstellar and how much is from the star.  Also, one can not distinguish whether there is more undetected matter from previous mass-ejection episodes, or whether some of the material was ejected while still on the main sequence.  Hence, for the total ejected mass, we will adopt a value of $0.09 \pm 0.05$  M$_\odot$ with a large  conservative estimate of the uncertainty in the  total mass loss for this star since it has begun to ascend the red giant branch.  This would correspond to a lifetime of between $0.8$ to $1.4 \times 10^5$ yr  in the red supergiant (RSG)  phase.  This provides a constraint on the models as discussed  below.

\section{Models}
\label{sec:models}
In this work,  spherical, non-rotating stellar evolution models  were calculated using two stellar evolution codes.  The first (henceforth referred to as the EG model) is the stellar evolution code originally developed by Eggleton (1971), but with updated nuclear reaction rates in an expanded network, and  the OPAL  opacities and EOS tables (Iglesias \& Rogers 1996).  
We have also made independent studies using the stellar evolution code MESA (Paxton et al.~2011; 2013).  The  MESA code utilizes the newer 2005 update of the
OPAL EOS tables (Rogers \& Nayfonov 2002) and the SCVH tables of
(Saumon, Chabrier  \& van Horn1995) to extend to lower temperatures and densities.  The latest version (Paxton et al.~2013) is particularly suited  to model massive stars.  Among the recent improvements, MESA  also includes
low-temperature opacities from either Ferguson et al.~(2005) or Freedman et al.~(2008) with updates to the molecular hydrogen pressure induced
opacity (Frommhold et al.~2010) and the ammonia opacity (Yurchenko et al.~2011).  These improvements to MESA  have an  effect on the deduced ages and evolution as we shall see.

Progenitor  models for $\alpha~ Orionis$ were constructed using the somewhat metal-rich progenitor composition $X= 0.70$, $Y = 0.28$, $Z = 0.024$ inferred (Lambert et al.~1984) from the surface [Fe/H] as discussed above.  Opacities and initial abundances were scaled from solar composition based upon our adopted metallicity.  With  the MESA code we  utilized the default abundances of Grevesse \& Sauval (1998).  For the EG code the Anders \& Grevesse (1989) abundances were employed.

 Models with the EG code typically utilized 300 radial mesh points held roughly constant in mass during the evolution.  Models generated with the MESA code included a variable mesh with up to about 3,000 radial zones.  The calculations were followed from the pre-collapse of an initial proto-stellar cloud through the completion of core carbon burning with the EG code.  Calculations with the MESA code were  run until silicon burning and were halted as the core became unstable to collapse.  Various mass-loss rates were analyzed  as described in \S 3.2, however, a normalized   (Reimers 1975) mass loss rate was ultimately adopted as the best choice.

Massive stars with M$\sim$ 10-25 M$_\odot$  are not  expected to have experienced much mass loss on the main sequence.  For a single isolated main sequence star of solar metallicity the only mass loss is via radiative winds.  Although there is some uncertainty in the mass loss rate,  previous studies have shown  (Woosley, Heger \& Weaver 2002; Heger et al.~2003) that no more than a few tenths of a solar mass are  ejected during the $\sim 10$ Myr main sequence lifetime.  This holds true even in models with rotation and magnetic fields (Heger, Woosley \& Spruit 2005).  
This is in contrast  to the much larger mass loss rate expected during the short RSG phase (Woosley, Heger \& Weaver 2002).   

Hence,  any mass loss that could have occurred on the main sequence would likely be small and would have dispersed into the interstellar medium by now.  Moreover, the small amount of expected mass loss should not  significantly alter the present observed properties of this star.  As a quick initial survey of models, therefore, we first ran hundreds  EG models  without mass loss.  We then   added mass loss as described below. The mass of the progenitor star for Betelgeuse is not known and estimates in the literature vary from 10 up to 25 M$_\odot$.  One goal of the present work, is to better determine a most likely mass for non-rotating models of this star.  Hence, progenitor models were constructed with masses ranging from $10 M_\odot$ to $75 M_\odot$.  Also,  the  mixing length parameter is not known and models were run with $\alpha$ ranging from 0.1 to 2.9.  A summary of the models run for these fits is given in Table \ref{tab:6}.

\begin{figure}[ht]
\plotone{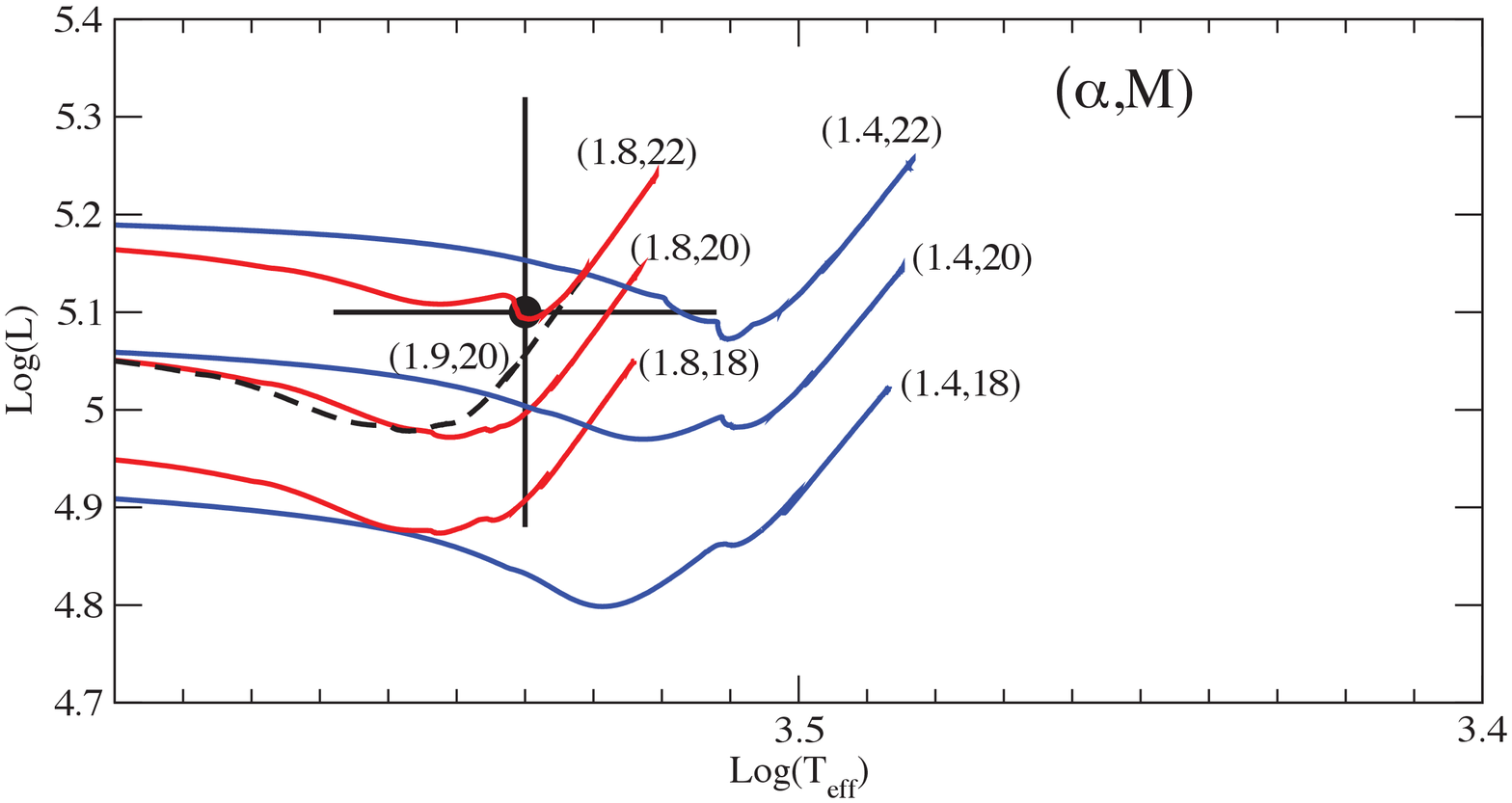}
\vskip .4 in
\caption{Illustration of the dependence of luminosity and temperature  on  progenitor mass M and  mixing length parameter $\alpha$. Lines show the HR diagram near the RSG for various MESA models  labeled by ($\alpha$,M).  The dashed line shows the best-fit model.}
\label{fig1}
\end{figure}

An illustration of the dependence of the observed luminosity and temperature on the mixing-length parameter $\alpha$ and the progenitor mass for models run with the MESA code is shown in Figure \ref{fig1}.
Here on can see that, for the most part,  the observed  present temperature fixes $\alpha$, while the observed luminosity (and requirement that the star has evolved past the first dredge-up) fixes the mass.  

A grid of over 500 models were run as summarized in Table \ref{tab:6}.  These models were  evaluated using a $\chi^2$ analysis.  The first analysis was  based upon a comparison of the EG models with the adopted constraints on  luminosity, radius, and surface temperature (ignoring mass loss).  In this step  $\chi^2$ was  determined from the simultaneous goodness of fit to $L, T$ and $R$.  Hence, we write:
\begin{equation}
\chi^2 = \sum_{i= L,T,R} \frac{(y_i^{obs} - y_i^{model})^2}{\sigma_i^2}~~,
\end{equation}
 where $y_i^{model}$ is the point along the evolutionary track for each model that minimizes the $\chi^2$ and $\sigma_i$ is the distance toward that point from the center to the surface of the 3-dimensional error ellipse.  The best fit from this calculation was found to be for masses in the  range M = $19^{+6}_{-2}$ M$_\odot$ and $\alpha=1.8^{+.7}_{-1.8}$.  
 
\begin{figure}[ht]
\plotone{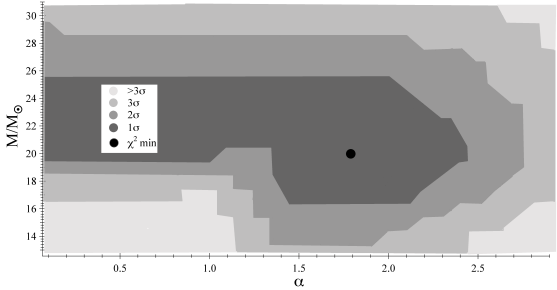}
\vskip .4 in
\caption{Contours of constant goodness of fit in the plane of progenitor  mass M versus mixing length parameter $\alpha$ for  EG models.  The shaded regions indicate the  1 $\sigma$ (66\%), 2 $\sigma$ (95\%) and 3 $\sigma$ (99.7\%) confidence limits as labeled.}
\label{fig2}
\end{figure}

 Figure \ref{fig2} shows contours in the mass versus mixing length parameter $\alpha$ plane.  Contours indicate the 1 $\sigma$ (66\%), 2 $\sigma$ (95\%) and 3 $\sigma$ (99.7\%) confidence limits.  The models with progenitor  masses from $14 M_\odot$ to $30 M_\odot$ were run again using an adopted  Reimers mass loss rate, as described below.  A mixing-length  parameter $\alpha = 1.8-1.9$ was chosen for these models, since $\alpha$ appears to have a shallow minimum around that value.  We note, however, that only an upper limit to $\alpha$ could be determined, and we will argue below based upon the observed surface convective velocities that a value of $\alpha = 1.4\pm 0.2$ is preferred near the surface.  
 Nevertheless, fitting the observed luminosity and temperature in particular require a value for $\alpha \sim 1.8-1.9$ as illustrated in Figures \ref{fig1} and \ref{fig2}.
 
 Models with mass loss were then evaluated again using both the the EG code and the MESA codes in a $\chi^2$ analysis based upon a comparison with not only the adopted luminosity, radius, and surface temperature, but also the  current adopted mass loss rate
 $\dot M$,  and 
the total ejected mass $M_{ej}$.  Since mass loss rate rises substantially as the models approach the base of the RGB, we integrate the ejected mass from that point to compare with the observed ejecta around the star.  In the these more constrained cases then  $\chi^2$ is   determined from:
\begin{equation}
\chi^2 = \sum_{i= L,T,R,\dot M, M_{ej}} \frac{(y_i^{obs} - y_i^{model})^2}{\sigma_i^2}~~.
\end{equation}

\begin{figure}[ht]
\plotone{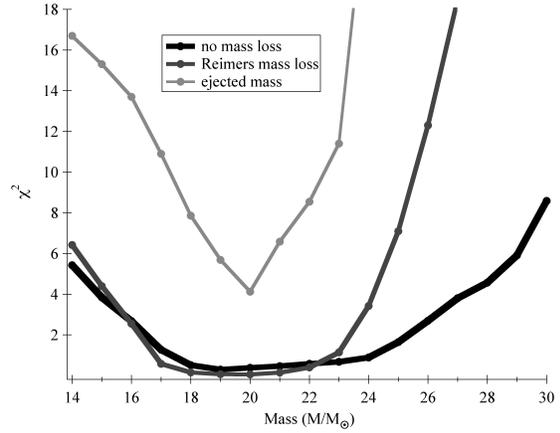}
\vskip .4 in
\caption{$\chi^2$ versus progenitor mass M for  as evaluated by three different $\chi^2$ analyses.  The black line shows the analysis for models evolved without mass loss and evaluated by comparing with observed luminosity, radius, and surface temperature.  The grey lines show the analyses for models evolved with a Reimers mass loss rate.  The dark grey line represents the analysis with the comparison with observed luminosity, radius, surface temperature, and mass loss rate, while the light grey line uses the same analysis but with total ejected mass as well. Notice how the minimum becomes better defined as more parameters are employed in the evaluation of $\chi^2$.}
\label{fig3}
\end{figure}

All of the  evaluations found  $\chi^2$ to be minimized for $M = 20$ M$_\odot$.  The best fits for both codes were  for  a progenitor mass of M = $20^{+5}_{-3}$ $M_\odot$.  Figure \ref{fig3} shows the comparison of the three $\chi^2$ analyses for $\chi^2$ versus mass based upon the EG models.   From this we deduce a best fit ($\sim~1\sigma$ C.L.) mass of   M = $20^{+5}_{-3}$ $M_\odot$  for both codes, and $\alpha=1.8^{+.7}_{-1.8}$  for the EG model, or $\alpha=1.9^{+.2}_{-0.6}$ with the MESA models.  The slightly larger value for $\alpha$ with the MESA code is due to the slightly higher opacities that cause the models to expand more in the RSG phase and have slightly cooler surface temperature for a given $\alpha$.

\begin{figure}[ht]
\plotone{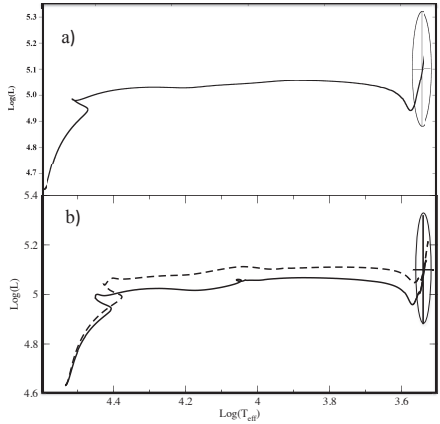}
\caption{HR diagrams for the best fit models according to the $\chi^2$ analysis.  The upper plot a) is computed with the modified EG code.  The lower track b) was computed with the MESA code. The dashed line on the lower plot shows the effect of convective overshoot with an overshoot parameter of $f=0.015$.   The error ellipse encloses the adopted uncertainty  in $T$ and $L$.}
\label{fig4}
\end{figure}

Figure \ref{fig4}a shows the HR diagram for the $20 M_\odot$ EG progenitor model with a Reimers mass loss rate, while \ref{fig4}b shows the track generated by the MESA code.  The tracks are nearly indistinguishable.  The error ellipse from the adopted constraints on $L$ and $T$ are also shown.  In both cases, the error ellipse encloses the track at the 1$\sigma$ level.  A summary of the "best fit" parameters deduced in this study is given in Table \ref{tab:6} and the implied observed properties are given in Table \ref{tab:7}.  The mass loss of this model is consistent with the observed values as we now discuss.  For the best fit models $\alpha$-Ori is currently on its  ascent as a red supergiant as expected and has not yet ignited  core carbon burning.  

Note that our conclusion that a non-rotating star of around 20 M$_\odot$ best fits the observations is consistent with the models of Meynet et al.~(2013).  In that paper it was 
also noted, however, that adding rotation, causes tracks to be  more luminous for a given
initial mass.  For example  in a rotating model
with an initial rotation with $v_{ini}/v_{crit} =0.4$, the current observed properties are   best fit with 
a progenitor mass of around 15 M$_\odot$.  This, however, corresponds to a rather high progenitor rotation rate.  The quantity  $v_{crit}$ is the maximum equatorial rotational velocity  such that the centrifugal force is exactly balanced by gravity.  

We also note, that our best-fit non-rotating mass and radius imply $\log{g} = -1.6$ and $R/M \sim 40$  (R$_\odot$/M$_\odot$).   This value of surface gravity is much smaller than the value $\log{g} = -0.5$ determined in the model atmospheres of (Lobel \& Dupree 2000).  The ratio R/M is similarly  a factor of two smaller than the value $82^{+13}_{-12}$ determined in the limb-darkening model of Neilson et al.~(2012).   If we allow for the fact that the model fits to stellar radius and mass are uncertain by about 25\% due to the uncertainties in the observed properties, then much of this discrepancy could be explained if the mass were near the lower range and the radius near the upper range of the uncertainty.  Just varying the mass alone with the observed radius would require M$\sim 12^{+5}_{-4}$ M$_\odot$ as pointed out in  Neilson, Lester \& Haubois (2012).  

Two remarks regarding these discrepancies are in order.  One is that these estimates are for the current mass of Betelgeuse and not the initial mass.  Possibly this  suggests an earlier epoch of  much more vigorous mass loss.  For example the mass estimate  of Neilson et al.~(2012) is consistent with the large mass loss in the rapidly rotating models of Meynet et al.~(2013).  Another possibility  is that  a larger radius   could also resolve these discrepancies.  This might, for example,  be due to the observed rather clumpy extended surface of  this star (e.g. Kervella et al.~2011; Chiavassa et al.~2011), i.e. the effective limb-darkening radius might be  larger than the spherical photospheric radius of the models.

\subsection{Convective overshoot}

We utilize  standard mixing-length theory treatment for convection.  The Ledoux criterion (i.e. including chemical inhomogeneities) is utilized to identify convective instabilities, and semi-convection is also included (i.e. mixing in regions that are Schwarzschild unstable though Ledoux stable) as described in the Mesa code (Paxton et al.~(2013). 

In a fully three-dimensional hydrodynamical treatment of convection
there can be   hydrodynamical mixing instabilities at convective boundaries. This is called  convective  overshoot.  As we shall see below, the outer region of Betelgeuse  consists of a rapidly  developing deep convective envelope extending deep into the star.  There is also convection  in the central helium-burning core. 
Because of this outer convective envelope, the observed properties of this star can be sensitive to the subtitles of convective overshoot at the boundary of the hydrogen-burning shell.  Hence, we have also made a study of the affects  of convective overshoot on the models derived here.

 In both codes this is accomplished via extra diffusive mixing at the boundaries of convective regions.  In the MESA code (Paxton et al.~2011; 2013) convective overshoot is parameterized according to the prescription of Herwig (2000).   That is, the MESA code sets an overshoot mixing diffusion coefficient
 \begin{equation}
 D_{OV} = D_{MLT} \exp{ -2 z/ (f H_P)}~~,
 \end{equation}
 where $D_{MLT}$ is a diffusion coefficient derived from mixing-length theory and $H_P = -P/(dP/dr)$ is the pressure scale height.  The free parameter $f$ denotes the fraction of the pressure scale height that extends a distance $z$ into to the radiative region.  A typical value for $f$ for AGB stars is $f \approx  0.015$  (Herwig 2000) with a maximum value consistent with observed giants of $f < 0.3$. We consider this range in the models.  

 However, as we shall see, although the addition of convective overshoot affects the location of the main sequence turn-off, it has very little effect on the giant branch.  Hence, the observed temperature and luminosity cannot be used to fix the convective overshoot parameter.  As we discuss below, however,  the observed surface abundances  place a constraint on the mixing parameter $f$.  Also as discussed below the age of this star is slightly decreased when convective overshoot is added.  Presumably this is due to changes in thermonuclear burning as material is mixed into radiative zones.
 
\subsection{Mass loss}

The total ejected mass since arriving at the base of the red giant branch (RGB)  is a small fraction of the total mass both observationally and in  the best fit models.       The mass loss rate increases considerably as the track approaches the base of the RGB.    Therefore, to better identify  the location of this star along its track we can compare  the observed accumulated ejected mass along with various integrated mass-loss rates since arriving at  the base RGB.  We note, however, that the mass loss begins slightly earlier in rotating models (Meynet et al.~2013).  

As shown in Table \ref{tab:5} our adopted  current observed mass loss rate is $2 \pm 1\times 10^{-6}$ M$_\odot$ $y^{-1}$ with  a total ejected mass of $0.09 \pm 0.05$ M$_\odot$ (Knapp \& Morris 1985; Glassgold \& Huggins 1986; Bowers \& Knapp 1987; Skinner \& Whitmore 1988; Mauron 1990; Marshall et al.~1992; Young et al.~1993; Huggins et al.~1994; Mauron et al.~1995; Guilain \& Mauron 1996; Harper et al.~2001; Plez et al.~2002; Ryde et al.~2006; Le Bertre et al.~2012; Humphreys 2013; Richards 2013).  

This would correspond to a lifetime of roughly between $3$ and $7\times 10^4$ yr in the red supergiant (RSG) phase.   However, one expects that the mass loss rate would have varied during the initial ascent from the base of the RGB.  Hence, we have considered  the various mass loss rates given below to integrate the total ejected mass from the  stellar models.

Realistically modeling the mass loss from Betelgeuse would be quite complicated.  It appears to be episodic (Humphreys 2013) and likely involves coupling with the magnetic field (Thirumalai, \& Heyl 2012) as well as the normal radiatively driven wind.  Indeed, there is evidence of a complex MHD bow shock around this star (Mackey et al.~2012; 2013; Mohamed,  Mackey \& Langer, 2012; 2013).  Nevertheless, for comparison of observations with models we have considered various parametrized mass-loss rates (Reimers 1975,1977; Lamers 1981; Nieuwenhuijzen \& de Jager 1990; Feast 1992; Salasnich et al.~1999).  For the Reimers (1975) rate,
\begin{equation}
\dot M = - 4 \times 10^{-13} \eta {L \over g R}~~{\rm M}_\odot {\rm yr}^{-1}~~,
\end{equation}
where $L,$ $R$, and $g$ are in solar units.  The observed rate requires a mass loss parameter of $\eta = 1.34 \pm 1$ for a $\sim20$ M$_\odot$ star of the adopted $L$ and $R$.  This value is not atypical for giants and is very close to the value inferred by Le Bertre et al.~(2012).  The Reimers (1977) rate,
\begin{eqnarray}
\lefteqn{\log{(-\dot M)} = } \nonumber \\
& & 1.50 \log{(L/L_\odot)} - \log{(M/M_\odot)} \nonumber \\
& & {} - 2.00 \log{(T_{eff})} - 4.74~~ 
\end{eqnarray}
implies a mass loss rate of $3.32 \times 10^{-6}$ M$_\odot$ $y^{-1}$ which is again consistent with observed rates and the Reimers (1975) value.
On the other hand, the Lamers (1981) rate
\begin{eqnarray}
\lefteqn{\log{(- \dot M)} = } \nonumber \\
& & 1.71 \log{(L/L_\odot)} - 0.99 \log{(M/M_\odot)} \nonumber \\
& & {} - 1.21 \log{(T_{eff} )} - 8.20 ~~,
\end{eqnarray}
gives a present mass loss of $8.80 \times 10^{-6}$ M$_\odot$ $y^{-1}$ which is higher by a factor of $\sim 4$ than  the adopted  current mass ejection rate.
The mass loss rate of de Jager et al.~(1988)
\begin{eqnarray}
\lefteqn{\log{(- \dot M)} = } \nonumber \\
& & 1.769 \log{(L/L_\odot)} \nonumber \\
& & {} - 1.676 \log{(T_{eff} )} - 8.158
\end{eqnarray}
would predict a mass loss rate of $ 8.40 \times 10^{-6}$ M$_\odot$ $y^{-1}$, which is also higher by a factor of $\sim 4$ relative to our adopted rate.  The rate from Salasnich et al.~(1999) 
\begin{equation}
\log{(- \dot M)} = -11.59 + 1.385  \log{(L/L_\odot)}
\end{equation} 
predicts a mass loss rate of $2.98 \times 10^{-5}$ M$_\odot$ $y^{-1}$, which is based upon the mass-loss pulsation-period relation of Feast (1992) 
\begin{equation}
\log{(\dot M)} = 1.32 \times \log{P} - 8.17 ~~.
\end{equation}
This implies a very high current mass loss rate of $1.96 \times 10^{-5}$ M$_\odot$ $y^{-1}$, (using the observed 420 day pulsation period).  

Based upon this comparison,   we adopt  the Reimers (1975)  rate with $\eta =   1.34 \pm 1$ as best representing this star.  We note, however,  that this total mass loss may only be a lower limit to the ejected mass along the RSG phase.  This is because not all of the ejected mass may be presently detectable if it was ejected sufficiently far in the past.

\subsection{HR Diagram}
Figures \ref{fig4}ab show  HR diagrams for the 20 M$_\odot$ modified EG model (a) and the MESA-code model (b).  
The HR diagrams are  quite similar.
The dashed line on the lower panel of Figure \ref{fig4} shows the HR diagram for a 20 M$_\odot$ model calculated with the MESA code with an overshoot parameter $f = 0.015$.
Here one can see that the main effect of the convective overshoot is to increase the luminosity of the main-sequence turn off and the base of the RGB.  It does not, however, affect the luminosity or temperature of the star as it moves up the giant branch.  Hence, there is little constraint on convective overshoot from the observed luminosity and temperature.  A value of $f = 0.3$, however, moves luminosity of the base of the RGB all the way up to the observed luminosity.  This much overshoot, however,  is inconsistent with the observed C and N abundances as we shall see below.

In both of our best-fit models, the first dredge-up occurs shortly after reaching the base of the RGB.  This is when the surface nitrogen is enriched.  This means that only models in which the current temperature and luminosity correspond to the ascent up the RSG phase can be consistent with this star.  

\subsection{Best Fit Model for the Present Star}

In spite of the large uncertainties, the adopted  mass-loss rate and ejected mass around this star can limit the possible present location of this star along its  evolutionary track if we accept that the star has only recently ascended the RSG phase as the best fit models imply.  Hence, we (somewhat arbitrarily) deduce the current age of the star from the amount of mass ejected from the base of the RGB.  We checked, however, and the deduced age and properties do not depend upon this assumption.
 
\begin{figure}[ht]
\plotone{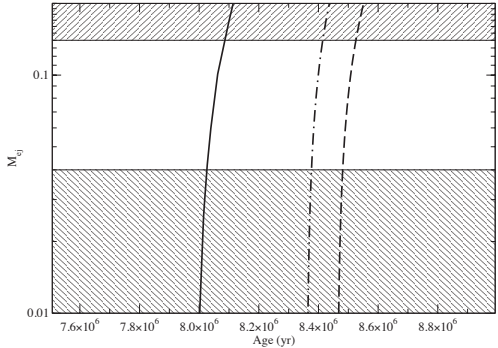}
\vskip .4 in
\caption{Thick solid line shows the total ejected mass $M_{ej}$ (in M$_\odot$) for the best-fit 20 M$_\odot$ progenitor model from the modified EG code as function of time.  The thick dashed line is from the MESA model.  The dot-dashed line is for  the MESA model with an overshoot parameter of $f = 0.015$.  Shaded regions  indicate the areas excluded by the adopted limits on the total ejected mass.}
\label{fig5}
\end{figure}

  Figure \ref{fig5} shows a comparison of the total mass ejected as a function of time  from the base of the RGB.  These curves are  based upon the Reimers (1975) rate with a mass-loss parameter of $\eta = 1.34$ for both the EG model (thick solid line) and the MESA (thick dashed  line) 20 M$_\odot$ progenitor models.   With our adopted parameters we find that by the time the star reached the main sequence turn off it has lost $0.1$ M$_\odot$ in both models, and by the time it reached the base of the base of the RGB   it had lost $0.3-0.4$ M$_\odot$.  It is possible, however,  that our adopted  Reimer's mass-loss parameter  overestimates the main sequence mass loss. 
   Winds during the MS phase are very different from the winds during  the RSG phase.  During the main sequence winds are mainly radiatively driven, while the RSG winds involve molecules, dust, etc.   Hence, these tracks represent upper and lower  limits for  the mass-loss evolution of this star.  
 
 The  shaded areas  indicate the excluded regions based upon our adopted uncertainty  the total ejected mass in Table \ref{tab:5}.  The MESA track indicates a  more advanced lifetime as it ascends the RGB.  The older age for the MESA models is mostly attributable to the increased opacity at low densities and temperatures in the MESA model.  This leads to lower luminosity, and hence, longer lifetimes.  
  In the context of these models, the amount of mass ejected corresponds to  a present age since the ZAMS of $\sim8.0$  Myr in the EG model and about  $8.5$ Myr in the MESA  model.  For both models the ages are about 0.1 Myr less when convective overshoot is included.  Here we adopt the ages in the MESA model without convective overshoot as the most realistic, but include the EG model results as an illustration of the uncertainty in this age estimate.

\subsubsection{Interior}
\begin{figure}[ht]
\plotone{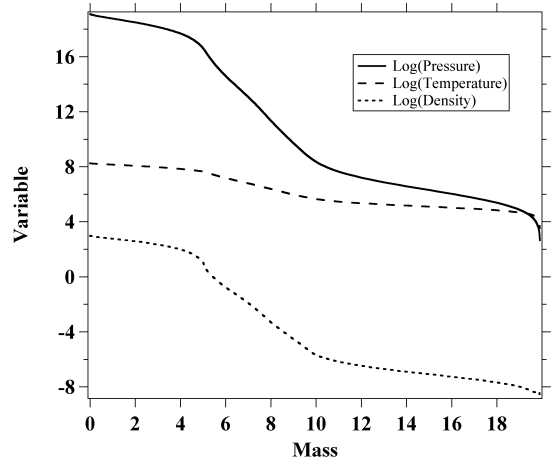}
\vskip .4 in
\caption{Thermodynamic state variables $\rho$, $T$, and $P$ as a function of mass for our best-fit 20 M$_\odot$ model.}
\label{fig6}
\end{figure}

Based upon our estimated  location in its ascent up the RSG phase, we now  examine the interior structure associated with this point in its evolution. Figure \ref{fig6} summarizes some of the interior thermodynamic properties.   The EG and MESA models give almost identical results for the interior thermodynamic properties.  In both models the star  is characterized at the present time by the presence of a developing  carbon-oxygen core up to  the bottom of the outer  helium core at $\sim 3-4$ M$_\odot$.

The  central density has risen to  $\sim10^3$ g cm$^{-3}$ with a central temperature of $\sim10^8$ K.  Outside of the developing C/O  core there is a region of steadily decreasing density in the outer helium core and hydrogen burning shell that extends to the outer envelope consisting of low density ($\sim10^{-6}$ g cm$^{-3}$) material.  The outer envelope   reaches to  90\% of the  mass  coordinate of the star.  This is followed by rapidly declining density and the  development of an outer surface convective zone.  

The best-fit MESA model left the main sequence  about $10^6$  yrs. ago, while for the EG model it was only about $3 \times 10^5$ years ago. Both models reached the base of the RGB about 40,000 years ago.   We followed the star through the final exhaustion of core helium burning in both codes,  followed by brief epochs of  core-carbon, neon, oxygen and silicon burning until core collapse and supernova an age  of 8.5 Myr since the ZAMS for the MESA code.   Our best guess is that the star will supernova in less than $\sim 100,000$ yrs (even longer in the EG model).  We note, however, that there error ellipse encompasses the entire track so that the star could be further along in its evolution.
The constraint that it has passed the first dredge-up, however, means that the star is ascending the RSG phase.  Our result is based upon mass loss from the base of the RGB is therefore a lower limit to how far it has evolved as a RSG.

\subsubsection{Composition}
\begin{figure}[ht]
\plotone{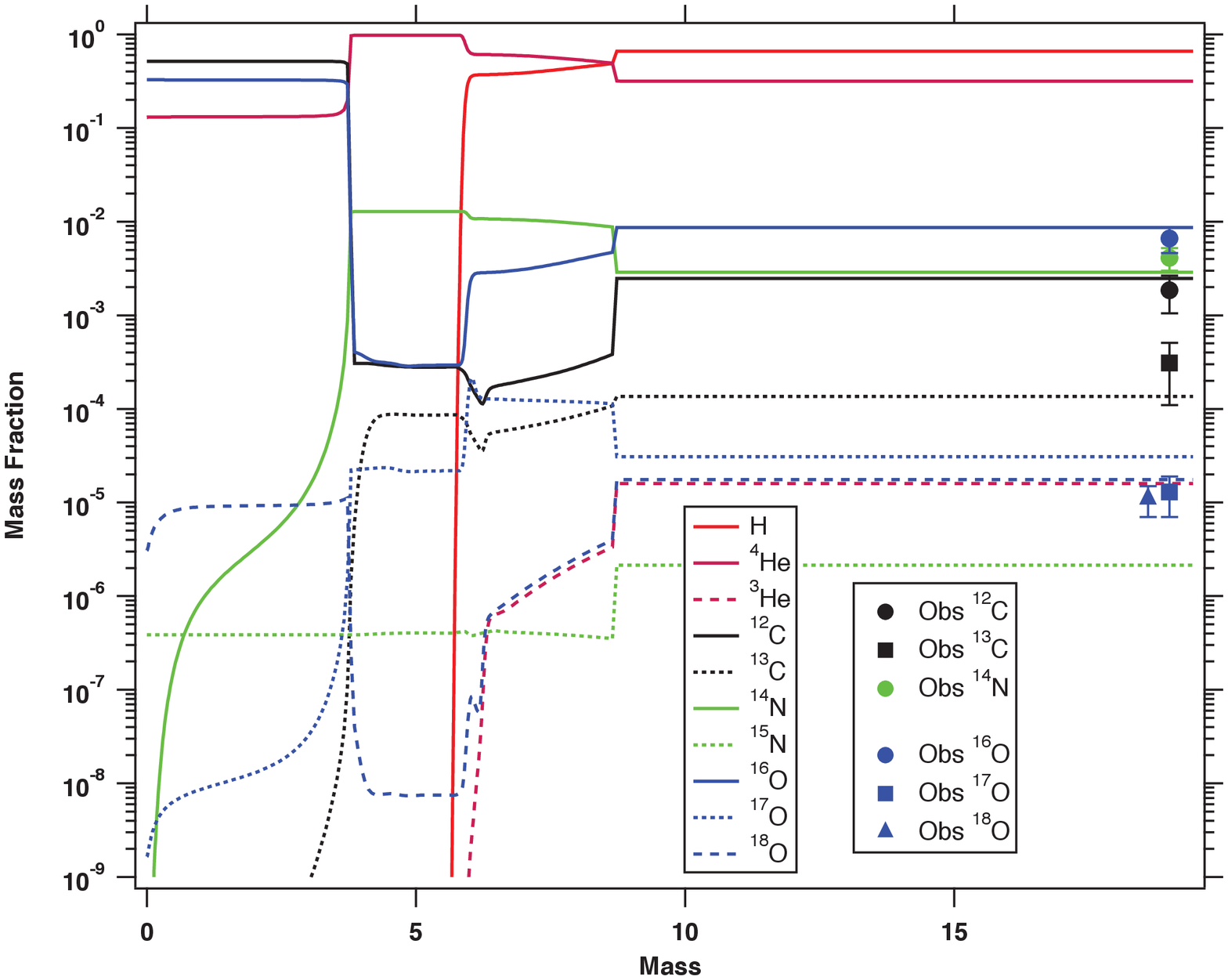}
\plotone{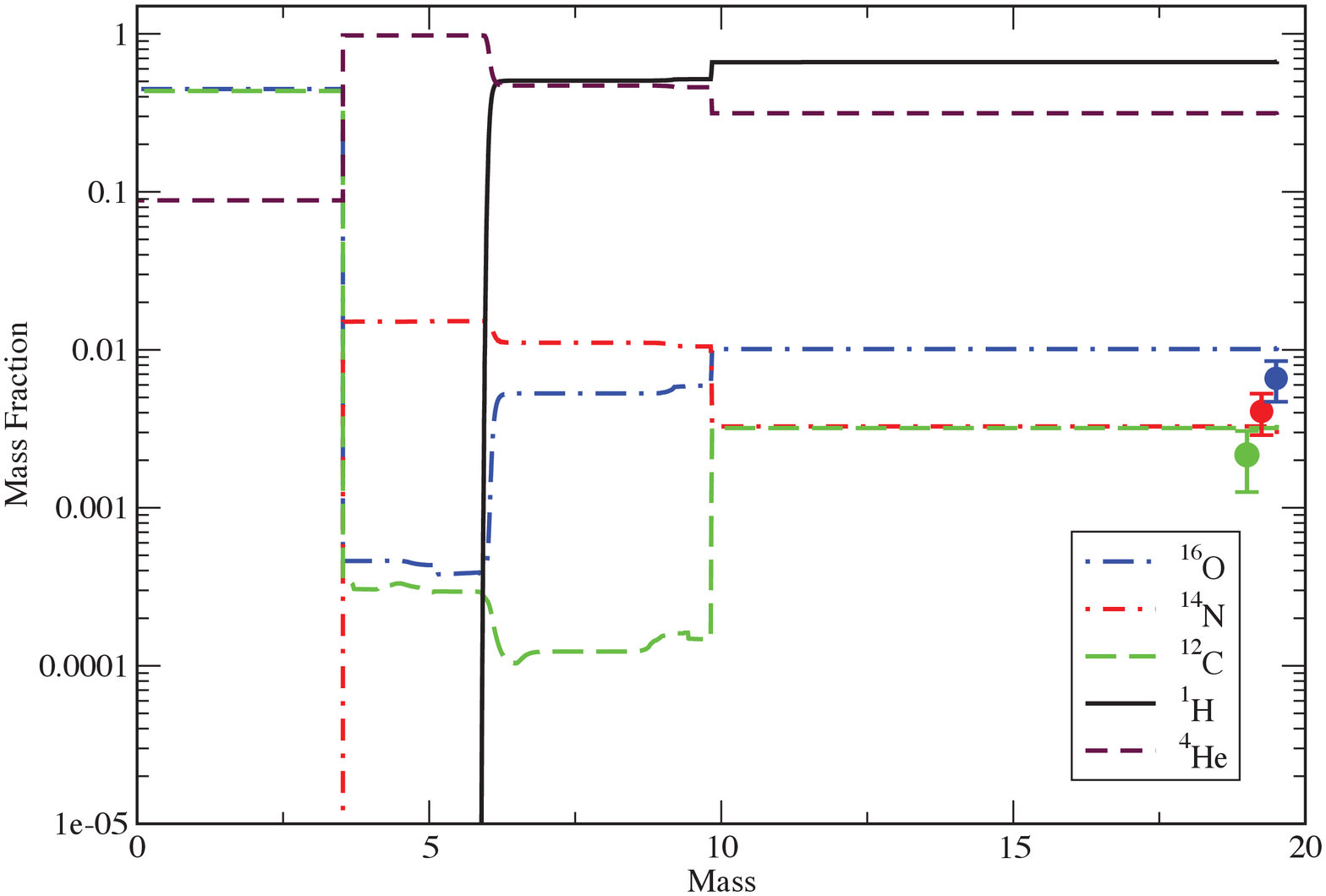}
\caption{Composite interior abundances as label a function of interior mass for the best fit 20 M$_\odot$ progenitor EG (upper panel ) and MESA (lower panel) models.  Points are surface abundances from Table \ref{tab:4}. (Color version online). }
\label{fig7}
\end{figure}

Figures \ref{fig7}a-\ref{fig7}b  show  composite plots of isotopic  abundances versus interior mass from the best fit EG and MESA 20 M$_\odot$ progenitor models.   These are  compared with the surface isotopic compositions determined  by Harris \& Lambert (1984) and  Lambert et al.~(1984).   
 The surface CNO surface abundances computed in both  models are quite consistent with the observations as shown by the points on the figures.  Hence, the models have correctly evolved the star through the first dredge up.  

The present-day interior composition were similar for the two models except that the interior carbon core is slightly larger in the MESA models due to the slightly  older lifetime.   Also, the convective core seems to be less efficiently mixed in the EG model than in the Mesa simulation.  We attribute this to the more sophisticated mixing treatment in the Mesa code (Paxton et al. 2013).  Nevertheless, in both models the C/O core has built up to a mass fraction of 40-50\% C and O and extends to about 3 M$_\odot$.  Above this, the He core extends to about 6 M$_\odot$, while the bottom of the outer convective envelope  is at 8 M$_\odot$ in the EG model, but has already descended  to $\sim10$ M$_\odot$ in the MESA model.      

\subsection{Convection and convective overshoot}
\begin{figure}[ht]
\plotone{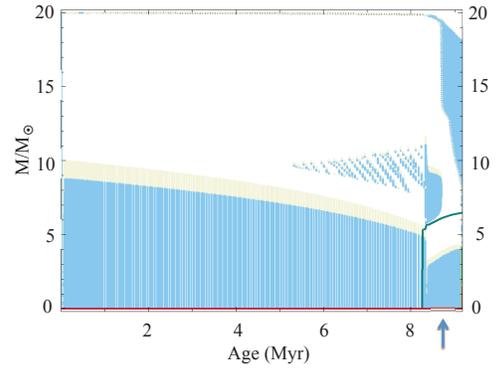}
\caption{Kippenhahn diagram showing evolution of convective zones for a  20 M$_\odot$ model computed with the MESA code with $\alpha = 1.8$ and a convective overshoot parameter of $f = 0.015$. Darker shading indicates  regions that are unstable to convection by the Ledoux criterion.  Lighter shading indicates the regions of convective overshoot.}
\label{fig8}
\end{figure}

The interior convective properties of the EG and MESA models were quite similar.    Figure \ref{fig8} shows a Kippenhahn diagram of the convective regions over the lifetime of a star for a 20 M$_\odot$ model with  $\alpha = 1.8$ and a convective overshoot parameter of $f = 0.015$.  An arrow at the bottom indicates 
our deduced present age for Betelgeuse.  Dark shaded regions are unstable to convection by the Ledoux criterion.  Lighter regions indicate the range of convective overshoot.
One can see  from this that the star presently has a rapidly developing outer convective envelope extending deep into the interior.   One can also see the recent onset of mass-loss for this star. 

\begin{figure}[ht]
\plotone{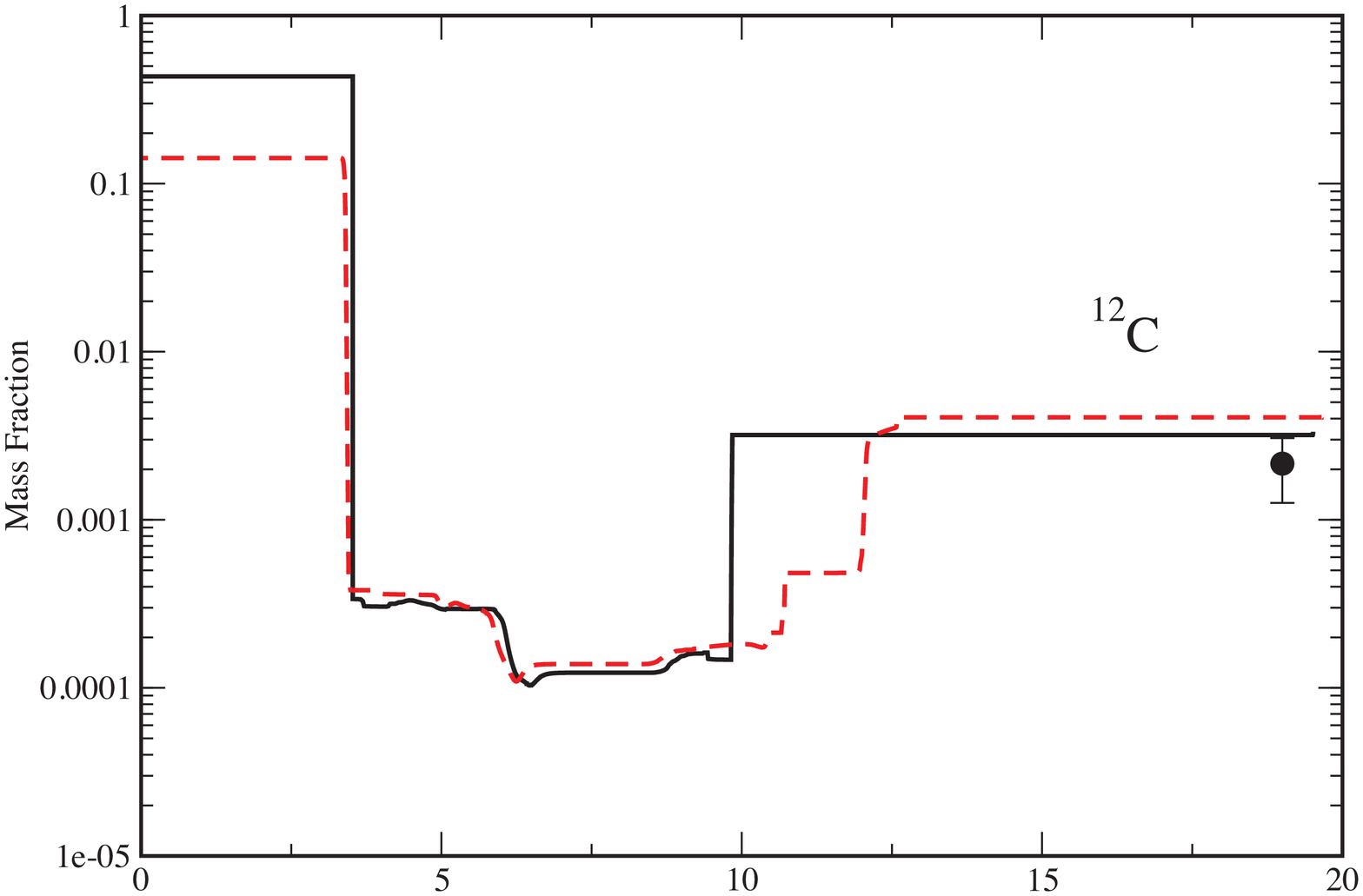}
\plotone{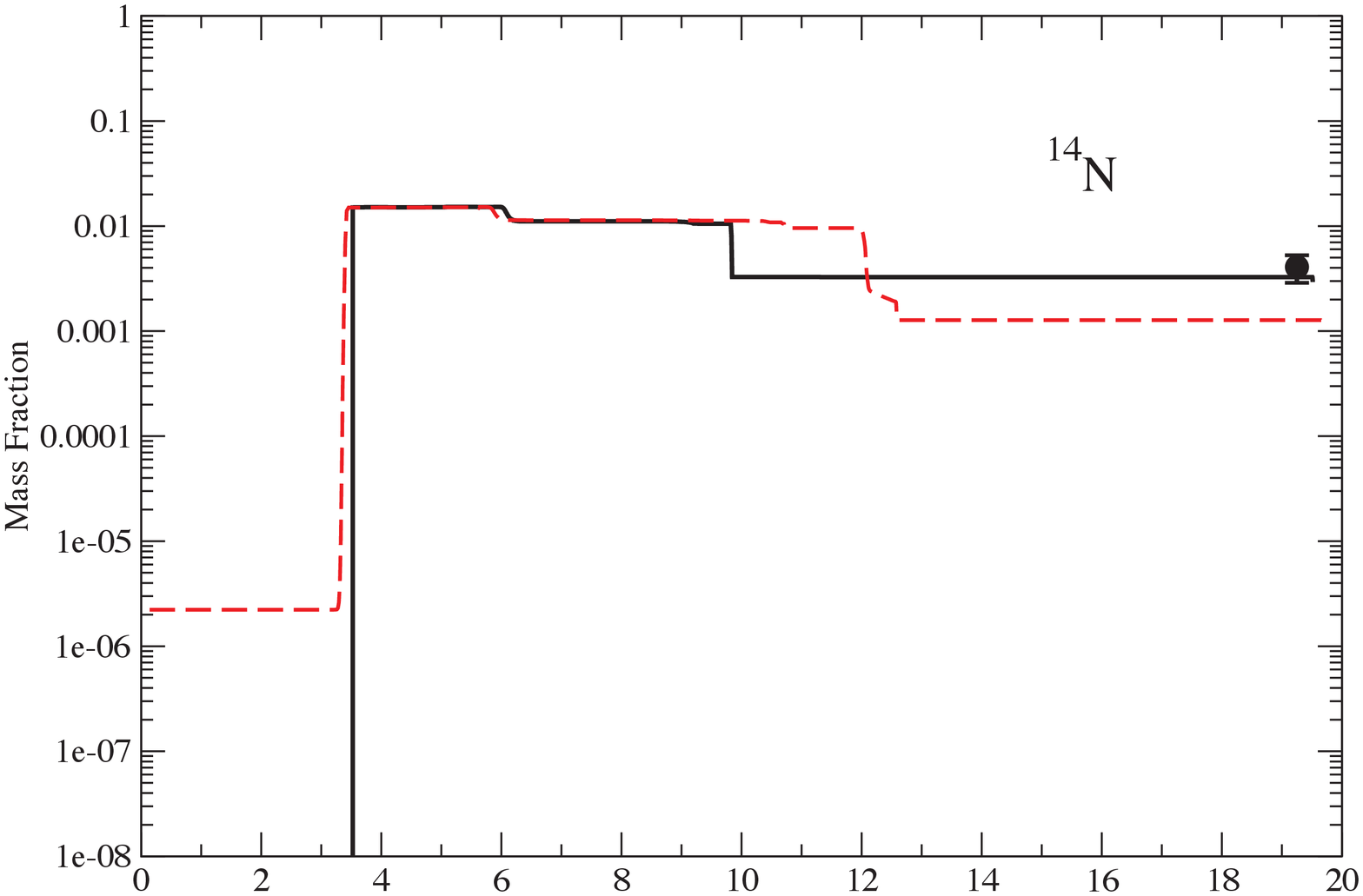}
\plotone{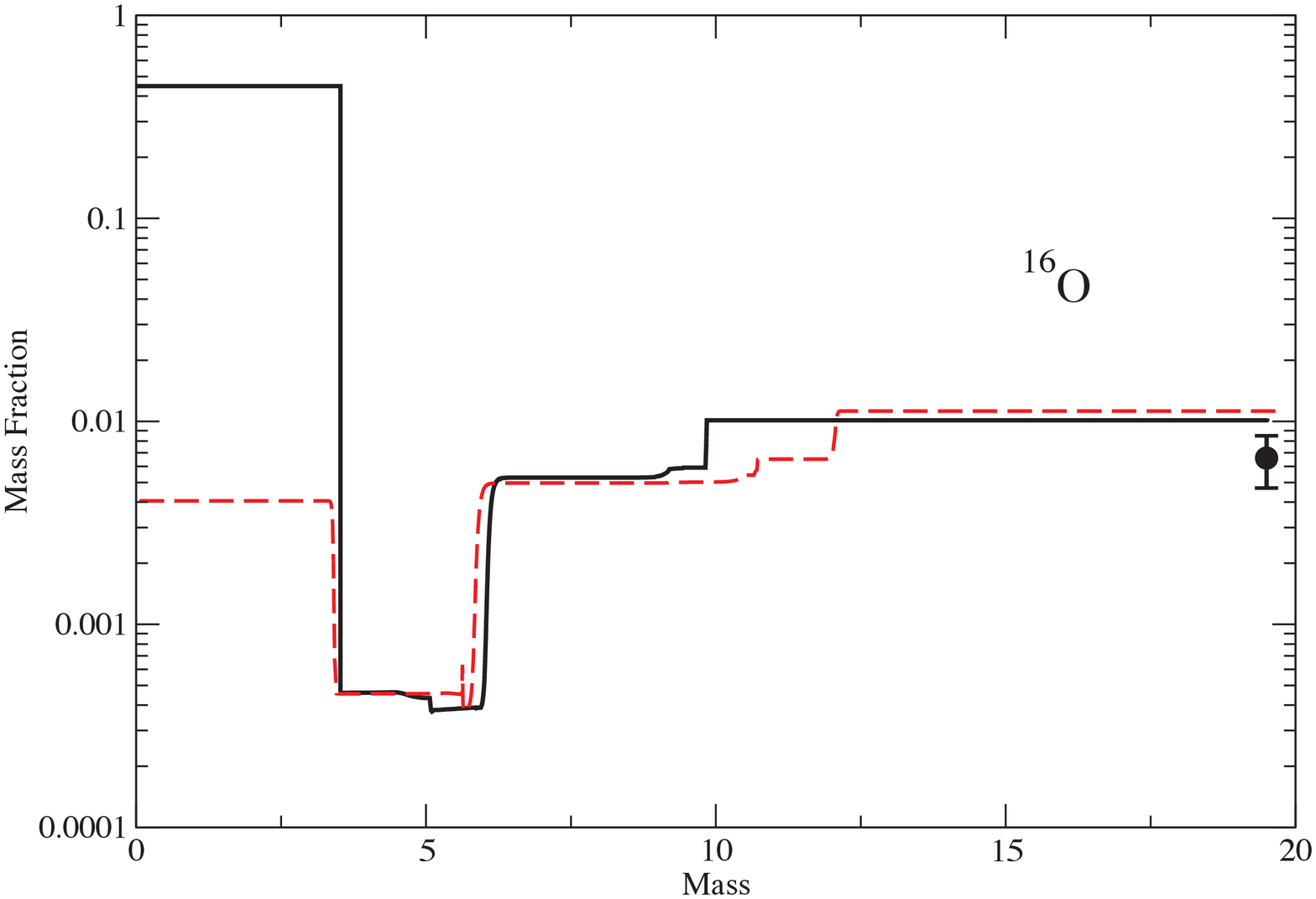}
\caption{Effects of convective overshoot on elemental composition.  Solid line shows interior C (upper panel), N (middle panel), and O (lower panel) abundances as a function of interior mass for the  20 M$_\odot$ progenitor MESA model.  Points are total elemental surface abundances from Table \ref{tab:4}. Dashed lines shows the effect of convective overshoot with mixing parameter $f = 0.015$.}
\label{fig9}
\end{figure}

The observed surface abundances, however, can be used to place constraints on convective overshoot  at the base of the outer convective envelope.  The three panels on Figure \ref{fig9} illustrate the effects of adding convective overshoot with $f = 0.015$ to the models.  One can see that the main effect of convective overshoot is to extend the bottom of the outer convective envelope from 10 to 12 M$_\odot$.  It also causes the surface abundances of C, N, and O to shift away from agreement, with the biggest discrepancy for $^{14}$N  which changes from agreement to a $3\sigma$ discrepancy.  The reason for these discrepancies is that convective overshoot tends to minimize the influence of the 1st dredge up.   Based upon this, we conclude that the overshoot parameter for the outer region is constrained to be $f < 0.10$ at the $2 \sigma$ confidence level and  are most consistent with $f = 0.0$.  Hence, we adopt $f=0.0$ for our best-fit models.  We note, however, that this constraint is not necessarily valid for the convective core where convection could behave quite differently (e.g. Viallet et al. 2015).

\subsection{Hot spots and convection}

It has been suggested (Buscher et al.~1990; Wilson et al.~1997, Freytag et al.~2002, Montarg\'es et al.~2015) that the occurrence of bright spots on the surface of Betelgeuse is the result of convective upwelling material at higher temperature.  It is possible to examine whether the occurrence of such features is consistent with simple mixing length theory in our  stellar evolution models.  That is, the distance $l_c$ over which a surface convective cell moves is characterized in mixing length theory by the pressure scale height.  
\begin{equation}
l_c = \alpha \biggl({P \over dP/dr}\biggr)
\end{equation}
The condition that a convective cell reach the photosphere is the
\begin{equation}
l_c \ge R - r
\end{equation}
where $r$ is the region from which the convective cell begins its upward motion.  The temperature $T_{hs}$ at which this cell appears on the surface as a hot spot is then given by
\begin{equation}
T_{hs} = T(r) + \int_r^R \biggl({dT \over dr}\biggr)_{ad} dr
\end{equation}
where the adiabatic temperature gradient is
\begin{equation}
 \biggl({dT \over dr} \biggr)_{ad} = \biggl(1 - {1 \over \gamma}\biggr) {T \over P} {dP \over dr}~~.
\end{equation}
 The change in luminosity due to the occurrence of such a spot is then
\begin{equation}
{\Delta L \over L} = \biggl({l_c \over R} \biggr)^2 \biggl({T_{hs} \over T}\biggr)^4
\end{equation}
where we have assumed that $l_c$ also characterizes the size of a convective cell  when it reaches the surface.  This seems justified by 3D simulations and images (Freytag, Steffen, \& Dorch 2002; Chiavassa et al.~2012; Kervella et al.~2012) that exhibit convective-cell profiles that are roughly spherical.  Indeed, a number of detailed three-dimensional numerical simulations
of deep convective envelopes have confirmed  the adequacy of mixing-length
theory except near the interior boundary layers (Chan \& Sofia
1987; Cattaneo et al.~1991; Kim et al.~1996).  In particular, they also  show 
approximately constant  upward and downward mean
velocities (Chan \& Sofia 1986), implying   a high
degree of coherence of the giant cells.

In Buscher et al.~(1990) a single bright feature was detected that contributed $\sim10-15$\% of the total observed flux.  In Wilson et al.~(1997) the observations were consistent with at least three bright spots contributing a total of 20\% to the total luminosity.  In their best fit model the hot spots were taken to have a Gaussian FWHM of 12.5 $mas$ corresponding to as much as a third of the total surface area in total or about 10\% of the surface per hot spot.  In Haubois et al.~(2009), they observed 2 hot spots attributing to a total of about 10\% of the total luminosity.  In Freytag et al.~(2002), the best fit for their radiation hydrodynamic model has luminosity variations of no more than 30\% for the total luminosity with numerous upwelling hot spots.  While their model supports the hot spot theory, and most parameters fit the data, they derive a mass of only $ 5-6 M _\odot$, which is inconsistent with observed luminosity and temperature for this star.  Further work by Dorch (2004) using magnetohydrodynamic modeling may improve on this inconsistency.  Here we point out that  a simple mixing-length  model is marginally adequate to explain the observed hot spots as we now describe. 

\begin{figure}[ht]
\plotone{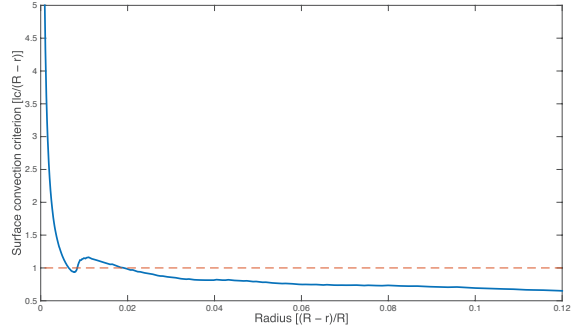}
\vskip .4 in
\caption{Surface convection criterion $l_c /(R - r)$ as a function of radius near the stellar surface a 20 M$_\odot$ model computed with the MESA code.}
\label{fig10}
\end{figure}

Figure \ref{fig10} shows the surface convection condition, $l_c/(R - r)$ as a function of interior radius for a 20 M$_\odot $ model with convective overshoot and $\alpha = 1.8$.  This shows that the typical size of a convective cell near the surface is less than about 2\% of the radius of the star.  Hence, the \t observed  surface hot spots  $\sim $ 10\% of the stellar disk would have to result from large  fluctuations in the size distribution.  The temperature change at the surface from our model is $\Delta T \approx 300$ K, which for a large fluctuation  in spot size would  correspond to a 20\% change in luminosity and  be  consistent with the observations.

We can also deduce a convective velocity from equating the work done in moving the convective cells to the kinetic energy in the bubbles.
\begin{equation}
v_c = {\biggl({\alpha k \over \mu m_H}\biggr) \biggl({T \beta \over g}\biggr)^{1/2} \biggl[\Delta
\biggl({dT \over dr}\biggr) \biggr]^{3/2}}~~,
\end{equation}
where $\Delta ({dT /dr})$ denotes the difference between the temperature gradient in the convective cell and that in the surroundings, $\alpha$ is the mixing length parameter, and $\beta = 1/2$ because we are measuring at the center of the convective cell.  We calculate a convective velocity at the surface of $v_c \sim$12 km s$^{-1}$.  Lobel (2001; 2003) deduced a value of $v_c = 9 \pm 1$ km s$^{-1}$.  The convective velocity has a dependence on $\alpha$ and can therefore serve as a constraint on $\alpha$.  The best fit models had $\alpha = 1.8-1.9$.  However, using the value of $v_c$ derived from Lobel \& Dupree (2001) and  Lobel (2003ab), would imply that $\alpha = 1.4 \pm 0.2$ near the surface.

An alternate explanation posed by Uitenbroek et al.~(1998) suggests the hot spots result from shock waves caused by pulsations of the stellar envelope.  They use Bowen's (1988) density stratification model calculated by Asida \& Tuchman (1995).  In this model, the stellar envelope pulsates, causing repetitive shock waves that create a density profile that is shallow with respect to the density predicted by hydrostatic equilibrium.  Future numerical work should be done to resolve which of the two scenarios is correct for $\alpha~ Orionis$.

\subsection{Periodic variability}

In addition to the random variability due to upwelling convective hot spots one also expects Betelgeuse to exhibit regular pulsations.  As  a massive star ascending the red giant, one expects $\alpha$-Ori to exhibit the regular radial pulsations associated with  a long period variable star.  Indeed, Betelgeuse is classified (Samus et al.~2011) as a semi-regular variable with a SRC sub-classification and a period of 2335 $d$ ($6.39~y$).   Analyzing the periodicity of this star is difficult, however, as  more than one  cycle appears to be operating simultaneously and intermittently.  In Dupree et al.~(1987) and  Smith, Patten, \& Goldberg (1989) a pulsational period of  420 days was detected that subsequently disappeared.  In Dempsey (2015) a pulsation period of 376 days was deduced using data in the AAVSO database.

It is worthwhile to analyze the  implied periodicity  from the best fit  model deduced here.  Although a detailed model is beyond the scope of the present paper, we can estimate the period to be expected from a one-zone  linear adiabatic wave analysis of radial oscillations (Cox \& Giuli 1968).  For a   surface shell on the exterior of a homogeneous star of radius $R$ and mass $M$ and a surface adiabatic equation of state index $\gamma$, the pulsation period $\Pi$ is just given by a linearization of surface hydrodynamic equations of motion for a radial perturbation $\delta R$ is
\begin{equation}
\frac{d^2(\delta R)}{dt^2} = -(-3 \gamma_{ad} - 4) \frac{GM}{R^3} \delta R~,
\end{equation}
which has a solution of simple harmonic motion corresponding to a period of,
\begin{equation}
\Pi = {2\pi \over \sqrt{(3 \gamma_{ad}  - 4)(4/3) \pi G \bar  \rho}} ~~,
\end{equation}
where $\bar \rho$ is the mean density of the star.  For our best fit model  the mean density is $\rho_0 = 10^{-7.2}$ g cm$^{-3}$ and the mean adiabatic index is $\gamma_{ad} = 1.5$.  This implies a pulsation period of $\Pi \approx 770$ $d$.  This seems reasonably close to the intermittent 420 d pulsation period, particularly given the simplicity  of the model and the fact that the observed period may be influenced by higher modes.  Clearly a more detailed nonlinear non-adiabatic analysis that would also determine the amplitude of the pulsations is warranted.  

We attempted a more realistic  radial pulsation calculation  for our model of $\alpha$ Orionis using both the pulsation code from Hansen \& Kawaler (1994) and the one in in the MESA code.  Although the calculated periods for certain harmonics agree  with the observed periodicity of $\alpha$ Orionis, the fundamental mode and first overtone could not be modeled well.  This problem may be  resolved by considering non-adiabatic pulsations coupled to the convection (Xiong, Deng, \& Cheng 1998).  Results of 3D simulations of RSG stars (Jacobs, Porter, \& Woodward 1999; Freytag et al.~2002)  and Betelgeuse in particular (Chiavassa et al.~2010; Freytag \& Chiavassa 2013) confirm the development of  large-scale granular convection that  generates hot spots.  Such convective motion should  also drive pulsations and should be analyzed in detail to explain the different pulsation frequencies for $\alpha$ Orionis.  Indeed, such large-scale convective motion may also contribute to the secondary periods of this star (Stothers 2010).

\section{Origin}
Based upon our inferred present age for this star, one can deduce the past positions of $\alpha~ Orionis$ using the {\it Hipparcos} measured proper motion of $\mu_{\alpha cos \delta} =24.95 \pm 0.08$  mas/y and $\mu_\delta = 9.56 \pm 0.15$ mas/y (Harper et al.~2008), combined with the measured radial velocity of $21.0 \pm .9$ km/s (Wilson et al.~1953).  The galactic coordinates for $\alpha$ Ori are (X,Y,Z)=(-121.8,-43.8,-20.4) and (U,V,W)=(21.4,-10.2,14.6).  Traced back ~10 Myr, this  gives (X,Y,Z)=(-339,59.9,-163.4).

The Orion OB 1a association has been considered a candidate for the origin of $\alpha~ Orionis$.  The association's age is $\sim 10$ Myr (Brown et al.~1994; Briceno et al.~2005), comparable with our predicted age for $\alpha~ Orionis$ of $\sim 8.5$  My.  The distance to the Orion OB 1a association is currently $336 \pm 16$ pc, measured using the mean distances to the subgroups (Brown et al.~1999).  However, the association is moving mainly radially  $\sim$28 km/s with negligible detected proper motion (Brown et al.~1999; de Zeeuw et al.~2000).  This motion traced back 10 My  gives a change in position of only $\sim 290$ pc, and is consistent with the implied  location of $\alpha~ Orionis$ at birth.  Recently, however, Bouy and Alves (2015) have discovered a new OB association that may be the origin of Betelgeuse. Nevertheless, one can conclude that $\alpha~ Orionis$ most likely did originate in the Orion nebula.  These results agree with a similar analysis done by Wing and Guinan (1997).

The age of $\alpha$ Ori also corresponds well with the age of the {\it Sco-Cen} subgroups.  However, the position of $\alpha ~Orionis$ 10 Myr ago is over 500 pc away from the position of the {\it Sco-Cen} cloud as determined by Mamajek, Lawson, \& Feigelson (2000).

\section{Future Supernova}
\begin{figure}[ht]
\plotone{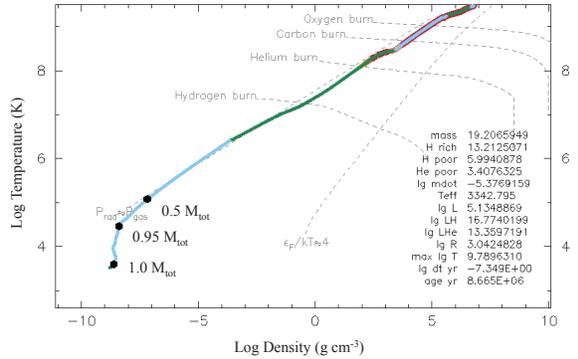}
\vskip .4 in
\caption{Interior temperature vs. density plot near the end of the lifetime of the best-fit   20 M$_\odot$ progenitor model obtained with the MESA code.}
\label{fig11}
\end{figure}

As for the future, $\alpha~ Orionis$ will continue burning He.   Eventually core C burning will begin, followed by core O burning and then core Si burning as it continues to increase in luminosity.  We estimate that in a little less than $10^5$ y, $\alpha~ Orionis$ will supernova, releasing $2.0 \times 10^{53}$ erg  in neutrinos along with $~2.0 \times 10^{51}$ erg in explosion kinetic energy (Smartt 2009) and leaving behind a neutron star of mass $\sim 1.5~M_\odot$.  
Figure \ref{fig11} shows the interior density and temperature profile for this star just prior to collapse.  At this point the star has evolved to an $\sim 1.5$ M$_\odot$ Si,Fe core and is about to collapse.

When this supernova explodes it will be closer than any known supernova observed to date, and about 19 times closer than Kepler's supernova.  Assuming it explodes as an average Type II supernova, the optical luminosity will be approximately -12.4, becoming brighter than the full moon.  The X-ray and $\gamma$-ray luminosities may be considerable, though not enough to penetrate the Earth's atmosphere.

  The interaction of such a supernova shock with the heliosphere has been studied in detail in Fields, Athanassiadou \& Johnson (2008).
  In that study it was demonstrated that unless a typical supernova occurs within a distance of 10 pc, the bow-shock compression of the heliosphere occurs at a distance beyond 1 AU.  Hence, for the adopted distance of 197 pc, the passing supernova remnant shock is not likely to directly deposit material on Earth.  Nevertheless, we can surmise some of the properties of the passing shock based upon a spherically symmetric Sedov-Taylor solution (Fields et al.~2008; Landau \& Lifshitz 1987).  The time for the arrival of the supernova shock front will be
  \begin{eqnarray}
  t &=& \beta^{-5/2} \sqrt{\frac{m_p n_{ISM}}{E_{SN}} } R_{SN}^{5/2} \\
  &= & 4.8 {\rm ~kyr} \biggl(\frac{10^{51}}{E_{SN}} \biggr)^{1/2}
  \biggl( \frac{n_{ISM}}{1 {\rm ~cm^{-3}}}\biggr)^{1/2} \biggl(\frac{R_{SN}}{10 {\rm ~pc}} \biggr)^{5/2},   \nonumber
  \end{eqnarray}
  where the   numerical factor  $\beta = 1.1517$ for an adiabatic index  of $\gamma = 5/3$, $m_p$ is the hydrogen mass, $n_{ISM}$ is the ambient interstellar medium density, $E_{SN}$ is the supernova explosion energy, and $R_{SN}$ is the distance to the supernova.
  For our adopted distance of 197 pc and explosion energy of $2 \times 10^{51}$ erg, we would expect the shock to arrive in about $6 \times 10^{6}$ yr.  The time scale for the passage of this supernova shock will be $> 1$ kyr.
As the shock arrives its velocity  will have diminished to:
  \begin{equation}
  v_{shock} = \frac{2}{5} \frac{R_{SN}}{t}~~ \approx 13 {\rm ~km~s^{-1}}~~,
  \end{equation}
  while the speed of the shocked material flowing behind the shock will be:
  \begin{equation}
  v_{SNR} = \frac{2}{\gamma + 1} v_{shock} \approx 10 {\rm ~km~s^{-1}}~~.
\end{equation}
The density of material behind the supernova shock will be $\approx 4 \rho_{ISM}$, while the total sum of ram pressure and thermal pressure behind the shock will be:
\begin{equation}
P_{SNR} = \frac{8 \beta^5}{25(\gamma - 1)}\frac{E_{SN}}{R_{SN}^3}  \approx 3.3 \times 10^{-9} {\rm ~ dyne~ cm^{-2}}~~.
\end{equation}
This is to be compared with the pressure of the solar wind for which 
\begin{equation}
v_{SW} \approx 450 {\rm ~km~s^{-1}}~~,
\end{equation}
\begin{eqnarray}
P_{SW} &= & \rho_{SW} v_{SW}^2 + P_{SW, thermal} \\
& \approx  & 2 \times 10^{-8} {\rm ~dyne ~ cm^{-2}} \biggl( \frac{{\rm 1 ~AU}}{R_{SW}} \biggr)^2~~~. \nonumber
\end{eqnarray}
A stagnation point will develop at the distance $R_{stag}$ at which the two pressures are equal giving
\begin{equation}
R_{stag} = \sqrt{\frac{P_{SW}}{P_{SNR}}} ~{\rm AU}~~,
\end{equation}
which for our adopted distance and energy implies:
$R_{stag} = 2.5$ AU, well beyond Earth.

\section{Conclusions} 

We have deduced quasi-static spherical  models for $\alpha~Orionis$ constrained by  presently known observable properties.  Table \ref{tab:7} summarizes a comparison between  the observable properties of this star with those from  the best-fit EG and MESA models.  As one can see, for the most part the models reproduce  the observed properties.  The fits to observations are optimized for  a 20 M$_\odot$ progenitor which is ascending the RSG phase and has passed through the first dredge-up phase near the base of the RGB.   It currently has a rapidly developing outer convective envelope.

A notable exception, however, to the agreement between predicted and observed properties is the surface gravity.  As discussed  above, a surface gravity of $\log(g) = -0.5$ was deduced from the model  atmospheres of Lobel \& Dupree (2000).  There is a  discrepancy of -.5  dex between that value and the value deduced from the  best-fit models.  Since $g$ scales as $M/R^2$ it is most sensitive to the radius and this discrepancy could be resolved if the effective radius of the surface gravity is much larger than the observed angular diameter.  The other possibility is that this star has undergone substantial mass loss, perhaps due to rapid rotation (Meynet et al.~2013).  

The mass loss, surface and core temperatures, and luminosity are  consistent with this star having relatively recently begun core helium burning.  As such, our best model is most consistent a mixing length parameter of $\alpha = 1.8-1.9 $,  and a present age of $t =8.5$ Myr.  The size and temperature of the convective up-flows are consistent with observed intermittent hot spots only if such spots correspond to large fluctuations in the typical surface convective cell size, and are probably due to more complicated magnetohydrodynamic evolution near the surface.  Also, the surface turbulent velocity seems to require a smaller mixing length parameter of $\alpha \sim 1.4$ near the surface.  The observed light-curve variability is  consistent with the derived mean density and equation of state for this star.   We estimate that this star will begin core carbon burning in less than $\sim 10^5 $ yr and will supernova shortly thereafter.  Although the supernova shock will arrive about 6 million years after the explosion, it is not expected that the supernova debris will penetrate the heliosphere closer than about 2.5 AU.

What is perhaps most needed now are good multidimensional turbulent models together with a nonlinear pulsation treatment to further probe the variability of this intriguing star during its current interesting  phase of evolution.

\acknowledgments
\nobreak
Work at Notre Dame is supported by DoE
Nuclear Theory grant number DE-FG02-95ER40934.  One of the authors (NQL) acknowledges support from the NSF Joint Institute for Nuclear Astrophysics.

\vfill\eject
\newpage

\begin{onecolumn}
\begin{deluxetable}{cccccc}
\tablewidth{0pt}
\tablecaption{Summary of Astronometric and Distance Data for Betelgeuse}
\tablewidth{0pt}
\tablehead{
\colhead{Ref.} & \colhead{p (mas)} & \colhead{distance (pc)} & \colhead{$v_{rad}$ (km s$^{-1})$} & \colhead{$\mu_{\alpha \cos{\delta}}$} & \colhead{$\mu_\delta$ } 
}
\startdata
Stanford (1933); Jones (1928) & &  & $20.7 \pm 0.4$ & & \\
Lambert et al.~(1984) & & $155$ & & & \\
Hipparcos  ESA (1997) & $7.63 \pm 1.64$ & $131^{+36}_{-23}$ &  & $27.33 \pm 2.30$ &$10.86 \pm 1.46$   \\
Tycho  ESA (1997) & $18.6 \pm 3.6$ & $54^{+13}_{-9}$ &  & &   \\
Famaey et al.~ (2005)) & &  & $21.91 \pm 0.51$ &  &   \\
Harper et al.~(2008) & $5.07 \pm 1.10$ & $197 \pm 45$ & & $24.95 \pm 0.08$ &$9.56 \pm 0.15$    \\ 

\bottomrule
Adopted & $5.07 \pm 1.10$ & $197 \pm 45$ & $21.91 \pm 0.51$ &  $24.95 \pm 0.08$ &$9.56 \pm 0.15$    \\
\bottomrule
\enddata
\label{tab:1}
\end{deluxetable}

%

\begin{deluxetable}{cccccc}
\tablewidth{0pt}
\tablecaption{Summary of Luminosity Temperature  Data for Betelgeuse} 
\tablewidth{0pt}
\tablehead{
\colhead{Ref.} & \colhead{V} & \colhead{$log(L/L_\odot)$} & \colhead{T$_{\rm eff}$ (K)} &  \\
}
\startdata
Lee (1970) &  $0.4$ & & & & \\
Wilson et al.~(1976) &  $0.7$ & & & & \\
Sinnott et al.~(1983) & $0.5 \pm 0.6$ & $4.67^{+0.44}_{-0.40}$ & & & \\
Lambert et al.~(1984) &  & &  $3800 \pm 100$ \\
Cheng et al.~(1986) &  $0.42$ & & &  \\
Gaustad et al.~(1986) & & & $3250^{+300}_{-120}$ & & \\
Mozurkewich et al.~(1991)& $0.5$ & &  \\
Dyck et al.~(1992) & $0.32 \pm 0.24$ & $4.74^{+0.31}_{-0.26}$ &$3520 \pm 85$ &  \\
Krisciunas et al.~(1992) & $0.43 \pm 0.14$ & $4.64^{+0.29}_{-0.27}$ & & & \\
Di Benedetto (1993)& & & $3620 \pm 90$ & & \\
Bester et al.~(1996) & & & $3075 \pm 125$ &  \\
Krisciunas et al.~(1996) & $0.59 \pm 0.24$ & $4.65^{+0.31}_{-0.23}$ & & & \\
Dyck et al.~(1996, 1998) & & &$3605 \pm 43$ &  \\
Wilson et al.~(1997) &  $0.64 \pm 0.20$ & $4.61^{+0.29}_{-0.25}$ &  \\
Perrin et al.~(2004) & & $4.80 \pm 0.19$ & $3641 \pm 53$  \\
Ryde et al.~(2006) & & & $3250 \pm 200$ & & \\
Harper et al.~(2008) & & $5.10 \pm 0.22$ & \\
Ohnaka et al.~(2011) & & & $3690 \pm 54$ & & \\ 
\bottomrule
Adopted &  $0.51^{+0.13}_{-0.19}$ & $5.10 \pm 0.22$ & $3500 \pm 200$ & \\
\hline
\enddata
\label{tab:2}
\end{deluxetable}

\begin{deluxetable}{cccccc}
\tablewidth{0pt}
\tablecaption{Summary of Angular Diameter/Radius  Data for Betelgeuse}
\tablewidth{0pt}
\tablehead{
\colhead{Ref.} & \colhead{Year Obs.} & \colhead{$\lambda (\mu m)$} & \colhead{$\Theta_{disk}^{\rm obs}(mas)$} &  \colhead{$\Theta_{disk}^{\rm corr}(mas) $} &  \colhead{R/R$_\odot$} 
}
\startdata
Michelson  (1921) &  1920 & 0.575 &  $47.0 \pm 4.7$ & $55 \pm 7$ & \\
Balega et al.~(1982) & 1978 & $0.405 - 0.715$   & $56 \pm 11$  & $57 \pm 7$ & $803^{+363}_{-234}$ \\
 & 1979 & $0.575 - 0.773$   & $56 \pm 6$  & $57 \pm 7$ & $803^{+363}_{-234}$ \\
Cheng et al.~(1986)   & & &   $42.1 \pm 1.1$ & & $593^{+182}_{-118}$ \\
Buscher et al.~(1990) & 1989 &  $0.633 - 0.710$ & $57 \pm 2$ & & $802^{+256}_{-165}$ \\
Mozurkewich et al.~(1991) & & & &  $49.4 \pm 0.2$ & $696^{+195}_{-126}$ \\
Dyck et al.~(1992) & & & &  $46.1 \pm 0.2 $ & $623^{+174}_{-113}$ \\
Wilson et al.~(1992) & 1991 & & & $54 \pm 2$ & $761^{+244}_{-158}$ \\
Bester et al.~(1996) & & & &  $56.6 \pm 1.$ & \\
Dyck et al.~(1996, 1998) & & & &  $44.2 \pm 0.2$ & \\
Burns et al.~(1997) & & & &  $51.1 \pm 1.5$ & \\
Tuthill et al.~(1997) & & & & $57 \pm 8$ & $803^{+415}_{-267}$ \\
Wilson et al.~(1997) & & & &  $58 \pm 2$ & $817^{+260}_{-170}$ \\
Weiner et al.~(2000) & 1999  &11.150 &  $ 54.7 \pm 0.3 $ & $ 55.2 \pm 0.5$& \\
Perrin et al.~(2004) & 1997 & 2.200&  $43.33 \pm 0.04$ & $43.76 \pm 0.12 $  & $620 \pm 124$  \\
Perrin et al.~(2004) &   & 11.15& $55.78 \pm 0.04$ &  $42.00 \pm 0.06$  & $620 \pm 124$  \\
Haubois et al.~(2009) & 2005 & 1.650 &  $44.3 \pm 0.1$ &  $45.01 \pm 0.12$ \\
Neilson et al.~(2011) & 2005 & 1.650 & - & $44.93 \pm 0.15$ & $955 \pm 217$ \\
Hernandez \& Chelli (2009) & 2006 &  $2.009 - 2.198$ & $42.57 \pm 0.02$ &  \\
Tatebe et al.~(2007) & & &  $48.4 \pm 1.4$ &  &\\
f et al.~(2009) & 2008 & $2.28 - 2.31$ & $43.19 \pm 0.03$ &  & \\
Townes et al.~(2009) &1993.83 & 11.150 &  $  56.0\pm   1.0$ & \\
& 1994.60  & 11.150 & $56.0  \pm 1.0  $ &  \\	  	
& 1999.875 & 11.150 & $ 54.9  \pm 0.3$ &  \\
& 2000.847  & 11.150 &  $53.4  \pm 0.6$ &  \\
& 2000.917  & 11.150 &  $55.8  \pm 0.9$ &  \\
& 2000.973  & 11.150 &  $54.8  \pm 1.0$ &  \\
& 2001.64  & 11.150 &  $53.4  \pm 0.6$ &  \\
& 2001.83  & 11.150 &  $52.9  \pm 0.4$ &  \\
& 2001.97  & 11.150 &  $52.7  \pm 0.7$ &  \\
& 2006.91  & 11.150 &  $48.4  \pm 1.4$ &  \\
& 2007.96  & 11.150 &  $50.0  \pm 1.0$ &  \\
& 2008.09  & 11.150 &  $49.0  \pm 1.5$ &  \\
& 2008.83  & 11.150 &  $47.0  \pm 2.0$ &  \\
& 2008.93 & 11.150 &  $ 47.0  \pm 1.0$ &  \\
& 2009.05  & 11.150 &  $48.0  \pm 1.0$ &  \\
& 2009.09  & 11.150 &  $48.0 \pm  2.0$ &  \\	 		
Ohnaka et al.~(2011) &2009 & $2.28 - 2.31$ & $42.05 \pm 0.05$ & $42.09 \pm 0.06$ & \\
\bottomrule 
Adopted & & &$55.64\pm 0.04$  & $41.9 \pm 0.06$ & $887 \pm 203$ \\
\hline
\enddata
 \label{tab:3}
 \end{deluxetable}


\begin{deluxetable}{ccccccc}
\tablewidth{0pt}
\tablecaption{Composition Data for Betelgeuse} 
\tablewidth{0pt}
\tablehead{
\colhead{Quantity} & \colhead{ } & \colhead{ }  & \colhead{Ref.} & \colhead{  } &  \colhead{ } & \colhead{ } 
}
\startdata
$[Fe/H] = +0.1$ &  &  &  Lambert et al.~(1984) \\
$X_\odot$ & $Y_\odot$ & $Z _\odot$   \\
0.71 & 0.27 &  0.020 & Anders \& Grevesse (1989)\\
$X$ & $Y$ & $Z$ \\
 0.70 & 0.28 & 0.024 & Corrected  for Betelgeuse (see text)  \\
\hline
$\epsilon$(C) & $\epsilon$(N) & $\epsilon$(O) &  & & \\
$8.41 \pm 0.15$ & $8.62 \pm 0.15$ & $8.77 \pm 0.15$&  Lambert et al.~(1984) & & \\
\hline
$^{12}$C/$^{13}$C & $^{16}$O/$^{17}$O  &$^{16}$O/$^{18}$O  & & & \\
$6 \pm 1$ & $525^{+250}_{-125}$ & $700^{+300}_{-175}$ & Harris \& Lambert (1984)& &  \\
\hline
\hline
Adopted Isotopic Mass Fractions \\
X($^{12}$C) & X($^{13}$C)  & X($^{14,15}$N) & \\
$1.85\pm 0.80 \times 10^{-3}$ & $0.31 \pm 0.20 \times 10^{-3}$ & $4.1 \pm 1.1 \times 10^{-3}$ &\\
X($^{16}$O) &X($^{17}$O)  &X($^{18}$O)  \\
$6.6 \pm 2.0 \times 10^{-3}$ & $1.3 \pm 0.6 \times 10^{-5}$ & $1.1 \pm 0.4 \times 10^{-5}$ \\
\hline
\hline
\enddata
\label{tab:4}
\end{deluxetable}

\begin{deluxetable}{ccccccc}
\tablewidth{0pt}
\tablecaption{Summary of Mass Loss and Variability Data for Betelgeuse} 
\tablewidth{0pt}
\tablehead{
\colhead{Ref.} & \colhead{$\dot{\rm M}$ (M$_\odot$ yr$^{-1}$)} & \colhead{M$_{ej}$  (M$_\odot$ )} & \colhead{(L$_{spot}/L$ -1)} & \colhead{T$_{spot}$ - T$_{eff}$ (K)} & \colhead{v$_{c}$ (km s$^{-1})$} 
}
\startdata
Knapp et al.~(1985) & $6 \times 10^{-7}$ & & & & \\
Glassgold et al.~(1986) & $4 \times 10^{-6}$ & & & & \\
Bowers et al.~(1987) & $9 \times 10^{-7}$ & & & & \\
Skinner et al.~(1988) & $1.5 \times 10^{-6}$ & & & & \\
Mauron (1990) & $4 \times 10^{-6}$ & & & & 3 \\
Marshall et al.~(1992) & $2 \times 10^{-7}$ & & & & \\
Young et al.~(1993) & $5.7 \times 10^{-7}$ & $0.042$ & & & \\
Huggins et al.~(1994) & $2. \times 10^{-6}$ & & & & \\
Mauron et al.~(1995) & $2 - 4 \times 10^{-6}$ & & & & \\
Guilain \& Mauron (1996) & $2 \times 10^{-6}$ & & & & \\
Noriega-Crespo et al.~(1997) & & $0.034 \pm 0.016$ & & & \\
Ryde et al.~(1999) & $2 \times 10^{-6}$ & & & & \\
Harper et al.~(2001) & $3.1 \pm 1.3 \times 10^{-6}$ & & & & \\
Plez et al.~(2002) & $2 \times 10^{-6}$ & & & & \\
Buscher et al.~(1990) & & & $10-15 \%$ & & \\
Wilson et al.~(1992) & & & $12 \pm 3 \%$ & & 6 \\
Tuthill et al.~(1997) & & & $11-23 \%$ & & \\
& & & $2-6 \%$ & & \\
Wilson et al.~(1997) & & & $13.1-15.0 \%$ & $600 $ & \\
& & & $4.0-7.3 \%$ & & \\
& & & $6.1-8.5 \%$ & & \\
Gilliland et al.~(1996) & & & & 200 & \\
Lobel et al.~(2001) & & & & & $12 \pm 1$ \\
Lobel (2003) & & & & & 12 \\
Le Bertre (2012) & $1.2 \times 10^{-6} $ & 0.086 & & &14  \\
\bottomrule
Adopted & $2 \pm 1 \times 10^{-6}$ & $.09 \pm 0.05$ &$12 \pm 12$\% & $400 \pm 200$ & $9 \pm 6$ \\
\hline
\enddata
\label{tab:5}
\end{deluxetable}

\begin{deluxetable}{ccccccc}
\tablewidth{0pt}
\tablecaption{Summary of Stellar Models Run in the Study} 
\tablewidth{0pt}
\tablehead{
\colhead{Code} & \colhead{$M_{prog}$ (M$_\odot$)} & \colhead{M$_{now}$ (M$_\odot$)} &  \colhead{$\alpha$} & \colhead{$\eta$ (MS to RGB)} & \colhead{$\eta $ (RGB)} & $f$ (Overshoot)   
}
\startdata
EG$^a$ & 10-75 &  10-75 & 0.1 - 2.9$^b$& 0.-1.34 & 0-1.34 & 0.0-0.3 \\
\hline
Best Fit EG & $20^{+5}_{-3}$ &   19.7&  $ 1.8^{+0.7}_{-1.8}$ & 1.34 & 1.34  & 0.0 \\
\bottomrule
\\
MESA$^b$ & 18-22 & 19.4 & 1.4 - 2.0  & 0.-1.34 &  1.34  & 0.0-0.3 \\
\hline
Best Fit MESA & $20^{+5}_{-3}$ &19.4  &  $ 1.9^{+0.2}_{-0.6}$ & 1.34 & 1.34  & 0.0 \\
\bottomrule
\enddata
\tablenotetext{a}{Over 510 EG models in increments of 1 M$_\odot$ from 10 to  30 M$_\odot$ , plus  50 and  75  M$_\odot$ and increments of 0.1 for $\alpha$.}
\tablenotetext{b}{Over 20 MESA models in M, $\alpha$, $\eta$ and $f$ near the best fit EG model.
 }
\label{tab:6}
\end{deluxetable}
\end{onecolumn}

\begin{deluxetable}{ccccccc}
\tablewidth{0pt}
\tablecaption{Summary of Fits to Observed Properties} 
\tablewidth{0pt}
\tablehead{
\colhead{Parameter} & \colhead{Observed} & \colhead{EG} & \colhead{MESA}   
}
\startdata
Age ($10^6$ yr) & - & 8.0 & 8.5 & \\
M (M$_\odot$) & - & 19.7 & 19.4 & \\
$M_{ej}$  (M$_\odot$)& $0.09 \pm 0.05$ &  $ 0.09$ & 0.09  &  \\
$\dot M$  ($10^{-6}$ M$_\odot$ yr$^{-1}$) & $2 \pm 1$ &  2.0 & 2.0 & \\
$\log{(L/L_\odot)}$ & $5.10 \pm 0.22$ & 4.97 & 4.99 &  \\
$T_{eff}$ (K)& $3500 \pm 200$ & 3630 & 3550 & \\
$R (R_\odot)$ & $887 \pm 203$ &  774 & 821 &  \\
$\log{(g) (cgs)}$ & -0.5 & -1.0 & -1.0 & \\
$R/M$ $(R_\odot/M_\odot)$ & 82$^{+13}_{-12}$ &  39 & 42 & \\
X($^{12}$C) ($10^{-3}$)& $1.85 \pm 0.80$ & 2.6  & 3.0 &  & \\
X($^{13}$C) ($10^{-3}$) & $0.31 \pm 0.20$  & 0.13  & - & \\
X($^{14}$N) ($10^{-3}$) & $4.1 \pm 1.1 $  &  2.3 & 3.7 & \\
X($^{15}$N)  ($10^{-3}$) & -  &   .0022 & - & \\
X($^{16}$O) ($10^{-3}$) & $6.6 \pm 2.0$ & 8.9  &  9.9 &  \\
X($^{17}$O) ($10^{-3}$) & $0.013 \pm 0.006$  & 0.0025 &  - & \\
X($^{18}$O) ($10^{-3}$) & $0.011 \pm 0.004$  & 0.0019 &  - &\\

\bottomrule

\enddata
 
\label{tab:7}
\end{deluxetable}
\end{document}